\documentclass[aps,prl,twocolumn,superscriptaddress,floatfix]{revtex4-2}

\usepackage{amsmath,amssymb,amsfonts}
\usepackage{graphicx}
\usepackage{bm}
\usepackage{xcolor}
\usepackage{hyperref}

\begin{document}

\title{Clausius' Second Law in Coupled Quantum Cavities:\\ Heat Flows, but Temperatures Need Not Equalize}

\author{Alejandro M. F. Rivas}
\email{alemfrivas@gmail.com}
\affiliation{Departamento de Física Teórica IFSC, Comisi\'on Nacional de Energ\'ia At\'omica (CNEA),
Av.~del Libertador 8250, Buenos Aires 1429, Argentina}
\affiliation{Consejo Nacional de Investigaciones Cient\'ificas y T\'ecnicas (CONICET), Argentina}

\author{Eduardo G. Vergini}
\affiliation{Departamento de Física Teórica IFSC, Comisi\'on Nacional de Energ\'ia At\'omica (CNEA),
Av.~del Libertador 8250, Buenos Aires 1429, Argentina}

\author{Leonardo Ermann}
\affiliation{Departamento de Física Teórica IFSC, Comisi\'on Nacional de Energ\'ia At\'omica (CNEA),
Av.~del Libertador 8250, Buenos Aires 1429, Argentina}
\affiliation{Consejo Nacional de Investigaciones Cient\'ificas y T\'ecnicas (CONICET), Argentina}

\author{Gabriel G. Carlo}
\affiliation{Departamento de Física Teórica IFSC, Comisi\'on Nacional de Energ\'ia At\'omica (CNEA),
Av.~del Libertador 8250, Buenos Aires 1429, Argentina}
\affiliation{Consejo Nacional de Investigaciones Cient\'ificas y T\'ecnicas (CONICET), Argentina}

\date{\today}

\begin{abstract}
Heat flows from hot to cold (Clausius' second law) until temperatures equalize (equilibration). We test both in two one-particle quantum cavities in contact. For thermal and coherent initial states heat flows hot to cold, but temperatures need not equalize, in chaotic and integrable cavities alike, despite strong entanglement. An analytical model explains why: heat moves in quanta between level pairs, and equalization needs resonant Bell pairs, ensured by exchange symmetry without eigenstate thermalization. Detuning or dark states break these pairs, and a few-mode contact cannot restore them.
\end{abstract}

\maketitle

\emph{Introduction.}---Two bodies at different temperatures placed in contact exchange heat, always from the hotter to the colder one, until they reach a common temperature. That heat never flows spontaneously the other way is Clausius' formulation of the second law of thermodynamics; that it keeps flowing until the temperatures are equal is a separate postulate, the approach to equilibrium, sometimes called the minus first law \cite{Brown2001}. Whether both follow from the reversible dynamics of the microscopic constituents is one of the oldest questions of statistical physics, and in quantum mechanics it remains open \cite{Gemmer2009,Gogolin2016}. When the two bodies start correlated, heat can flow from cold to hot \cite{Partovi2008,Jennings2010}, as has been observed experimentally \cite{Micadei2019}; the second law requires uncorrelated initial states, the quantum analogue of Boltzmann's \emph{Stosszahlansatz}. We start from product states, and ask whether the flow and its end at equal temperatures emerge from the dynamics.

The usual expectation rests on a chain of ideas that starts with symmetry. A symmetry of the Hamiltonian implies a conserved quantity; enough conserved quantities make a system integrable; their absence leads to chaos. A classical chaotic billiard such as the stadium is ergodic and mixing \cite{Bunimovich1979}, and ergodicity is the traditional foundation of statistical mechanics. Its quantum counterparts, quantum ergodicity of eigenfunctions \cite{Shnirelman1974}, random-matrix spectral statistics \cite{Bohigas1984}, and eigenstate thermalization \cite{Deutsch1991,Srednicki1994,Rigol2008,Dalessio2016}, explain why chaotic quantum systems look thermal \cite{Borgonovi2016}. One therefore expects chaos to favor thermalization and symmetries to hinder it; in integrable systems, conserved quantities indeed prevent relaxation to Gibbs states \cite{Rigol2007}.

In practice, thermalization is diagnosed by the saturation of the entanglement entropy \cite{Miller1999,Kaufman2016,Bergamasco2017,Bergamasco2019}, or by the approach to Gibbs states, which however follow from maximizing the entropy of the whole only when one part is much larger than the other and acts as a bath \cite{Goldstein2006,Popescu2006}. When the two parts have the same size there is no reason to expect them: for two identical systems in contact, Ponomarev \emph{et al.} found that the populations reach the arithmetic mean of the initial ones at weak coupling, and Gibbs-like populations when the coupling spreads the populations uniformly within each energy shell \cite{Ponomarev2011,Ponomarev2012}. For two chaotic subsystems the eigenstates go from product-like to random as the coupling grows \cite{Lakshminarayan2016,Tomsovic2018,Pulikkottil2023}. With no bath, the test that remains for two bodies of the same size is the condition of thermal equilibrium itself: equal temperatures.

Here we study two cavities with one particle each, in contact through a window. We model them as quantum billiards, realized in microwave resonators \cite{Stockmann1990} and quantum dots \cite{Marcus1992}. A single particle in a billiard can be assigned a temperature through equipartition, and a pressure, related by the ideal gas law; interpreting the wave function as a probability density analogous to that of an ideal gas gives these quantities their thermodynamic meaning \cite{Rivas2025}; the two particles interact precisely through these pressures. We find that, for thermal and coherent initial states, heat flows in the direction Clausius requires, but does not always stop at equal temperatures. An analytical model explains both results. Its key is energy quantization: heat is exchanged in discrete quanta, one pair of levels at a time, and through a local contact complete equalization requires these pairs to form Bell states. Exchange symmetry provides them; a detuning, or the dark states of cavities with additive spectra, prevents them; and a contact with few modes cannot restore them, even when made stronger. Entanglement, chaos and eigenstate thermalization do not decide the outcome. The analytical model is derived in the End Matter; numerical details and robustness checks are in the Supplemental Material (SM).

\emph{Model and temperature.}---With $\hbar=1$ and $m=1/2$, the Hamiltonian is $H=H_1+H_2+\lambda H_{\text{int}}$, where $H_{1,2}=-\nabla^2_{1,2}$ describe the isolated cavities with Dirichlet conditions, with eigenstates $\phi_i$ and energies $\varepsilon_i$, primed for cavity 2 when the cavities differ. Because the wave functions vanish on the walls, the particles can only interact through the normal derivatives of their wave functions at the window, and $|\partial_n\psi(y)|^2$ is the local pressure on the wall \cite{Barnett2000,Rivas2025}. We take the interaction to be the product of the two pressures at each point of the window,
\begin{equation}
\begin{split}
H_{\text{int}}&=\int_w dy\;\hat P_1(y)\,\hat P_2(y),\\
\langle i|\hat P(y)|k\rangle&=\partial_n\phi_i(y)\,\partial_n\phi_k(y),
\end{split}
\label{eq:hintP}
\end{equation}
in a basis of $N_1=80$ levels per cavity obtained with the scaling method \cite{Vergini1995}. We consider a desymmetrized Bunimovich stadium, which is chaotic, and a quarter circle (hereafter, the circle), a square and rectangles, which are integrable. The second cavity is detuned by changing its shape parameter $l_s$ (the straight segment of the stadium) from $l_{s,1}=0.5$ to $l_{s,2}$. Unless stated otherwise $\lambda=10^{-5}$. Equation~(\ref{eq:hintP}) is an effective model of a local contact, a movable wall as in the classical adiabatic piston \cite{Lieb1999,Gruber1999,Kestemont2000}; like any contact interaction in two dimensions it needs an ultraviolet cutoff, here the basis; a smooth cutoff converged in the basis size gives the same conclusions (SM, Secs.~S1, S2 and S9).

The initial state is a product of coherent states, cold in cavity 1 and hot in cavity 2. The temperature of each cavity is defined, in or out of equilibrium, by equipartition \cite{Rivas2025,CasasVazquez2003,Puglisi2017}, $T_i(t)=\mathrm{Tr}[\rho_i(t)\hat T_i]$, where $\rho_i$ is the reduced density matrix of cavity $i$ and the temperature operator $\hat T_i$ is its kinetic energy in units of $k_B$ ($k_B=1$). It is operational: the cavities are not reservoirs and their states need not be of Gibbs form, although fitting the populations to one gives the same conclusions. In a finite isolated system the temperatures keep fluctuating, so they can only become equal \emph{within their fluctuations}. We compare the late-time gap $\Delta T=|\langle T_1\rangle-\langle T_2\rangle|$ with the amplitude $\sigma$ of the fluctuations, and say that the temperatures equalize when $\Delta T/\sigma<1$. Thermal equilibrium would give equal temperatures, and so does the microcanonical ensemble of the composite system, for all the pairs studied. To quantify how much of the imbalance survives we use $\eta=\Delta T/(\langle T_1\rangle+\langle T_2\rangle)$ and the transferred fraction $1-\eta/\eta_0$, with $\eta_0$ its initial value. Both $\Delta T$ and $\sigma$ can be predicted from the eigenstates without solving the dynamics (SM, Sec.~S3).

\begin{figure}[t]
\includegraphics[width=\columnwidth]{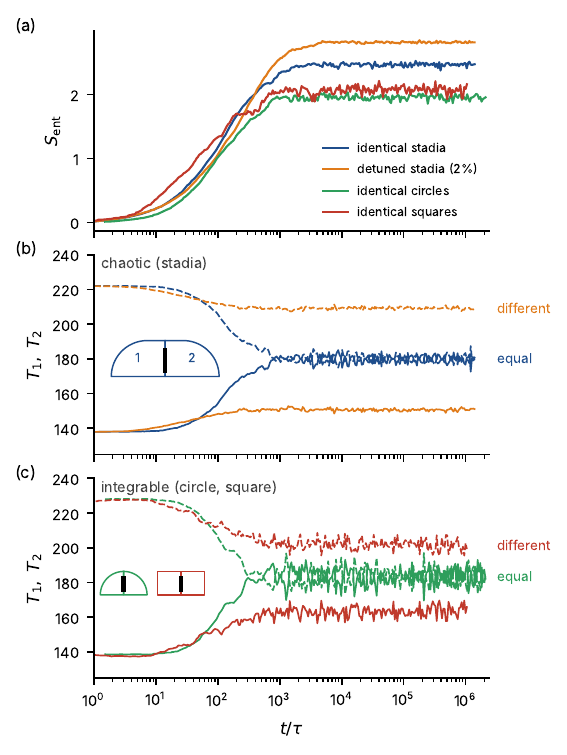}
\caption{\textbf{Entanglement saturates, temperatures need not equalize.} (a) Entanglement entropy $S_{\text{ent}}=-\mathrm{Tr}\,\rho_1\ln\rho_1$. (b), (c) Temperatures of the cold (solid) and hot (dashed) cavity, colors as in (a): (b) chaotic cavities, identical stadia and stadia detuned by $2\%$ ($l_{s,2}=0.51$); (c) integrable cavities, identical circles and identical squares. Insets: the cavities, cold (1) and hot (2), and the window (thick segment).}
\label{fig:entropy}
\end{figure}

\emph{Entanglement saturates, temperatures need not equalize.}---Figure~\ref{fig:entropy}(a) shows the entanglement entropy for four pairs of cavities; in all of them it grows and saturates. The circles and the squares, both integrable, saturate at nearly the same value; among the chaotic stadia, the detuned pair saturates higher than the identical one, near the value for random phases: dephasing entangles them without moving energy (SM, Sec.~S4).

The temperatures tell a different story [Figs.~\ref{fig:entropy}(b),(c)]. In all four cases energy flows from the hot to the cold cavity, but only two pairs end at a common temperature: the identical stadia and the identical circles. The detuned stadia, although chaotic and more entangled, keep two different temperatures, and so do the squares, although they are as entangled as the circles. The saturation of the entanglement entropy does not imply equal temperatures, and neither chaos nor integrability of the cavities decides the outcome. The statistics of the two-particle levels does not decide either: the coupled detuned stadia have nearly random-matrix level statistics and do not equalize, whereas a much smaller detuning, with nearly Poisson statistics, comes close to equalization (SM, Sec.~S4).

\begin{figure*}[t]
\includegraphics[width=0.95\textwidth]{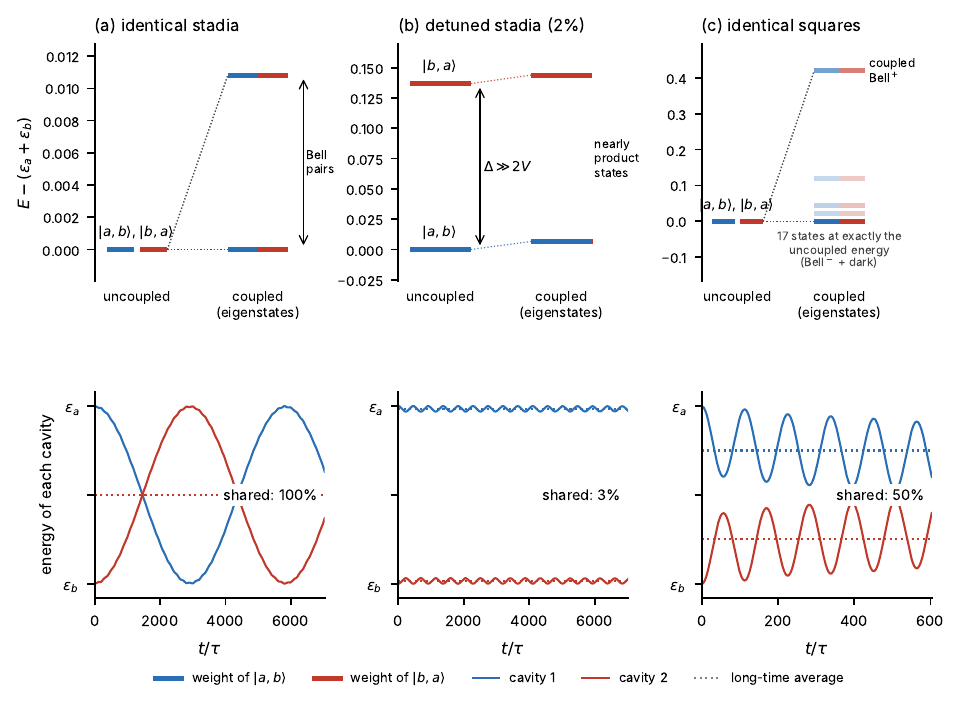}
\caption{\textbf{Energy is shared channel by channel.} Each pair of levels $\{a,b\}$ forms a two-state system, $|a,b\rangle$ and $|b,a\rangle$ (level $a$ in cavity 1, $b$ in cavity 2, and the reverse); going from one to the other carries $\varepsilon_a-\varepsilon_b$ across the window. Top: energy levels. Left, the uncoupled $|a,b\rangle$ (blue) and $|b,a\rangle$ (red); right, the eigenstates of $H$, colored by their weights on each. Bottom: evolution from $|a,b\rangle$: energy of each cavity (solid) and its long-time average (dotted). (a) Identical stadia, channel $\{1,31\}$: $|a,b\rangle$ and $|b,a\rangle$ are degenerate and the coupling turns them into two Bell states; the energy is equally shared on average. (b) Stadia detuned by $2\%$, same channel: split by $\Delta\simeq9\,V$, each eigenstate remains almost purely one of them, and only $3\%$ is shared. (c) Identical squares, channel $\{2,21\}$, one of $14$ channels of the same total energy: $75\%$ of its weight lies in $17$ eigenstates that stay at the uncoupled energy (the Bell$^-$ state and dark combinations), and only half is shared. $\lambda=10^{-5}$.}
\label{fig:schematic}
\end{figure*}

\emph{Quantization, symmetry and channels.}---The analytical model rests on energy quantization. Classically, two bodies in contact can exchange any amount of energy. Quantum mechanically, without coupling the eigenstates are products $|a,b\rangle$ with energy $\varepsilon_a+\varepsilon_b$, and a weak coupling conserves this energy to within its own strength: cavity 1 can give up the quantum $\varepsilon_a-\varepsilon_b$ only if cavity 2 can absorb the same quantum. Each pair of levels $\{a,b\}$ defines a channel between $|a,b\rangle$ and $|b,a\rangle$, and heat exchange becomes a resonance problem: a channel transfers energy only if its two states are degenerate, or closer in energy than their coupling $V$ [Eqs.~(\ref{eq:A1})--(\ref{eq:A4})]. Figure~\ref{fig:schematic} shows the three situations we find.

When the cavities are identical, exchanging the particles is a symmetry of $H$, and the exchange parity is conserved. If the spectrum has no exact degeneracies, every eigenstate is then symmetric or antisymmetric, holds the same energy in each cavity, and has $\langle n|\hat T_1-\hat T_2|n\rangle=0$ [Eq.~(\ref{eq:A7})]. Since at long times only these diagonal elements survive, the two cavities reach a common temperature for any initial state, whatever the coupling, as long as it respects the symmetry. In the channel picture, $|a,b\rangle$ and $|b,a\rangle$ are degenerate, any coupling mixes them into the Bell states $(|a,b\rangle\pm|b,a\rangle)/\sqrt2$, and the energy oscillates back and forth across the window, equally shared on average [Fig.~\ref{fig:schematic}(a)]. The same holds for five initial states and three other couplings (SM, Secs.~S2 and S6). This equalization does not erase the memory of the initial state: each cavity ends with the average of the two initial distributions [Eq.~(\ref{eq:A8})]. Here a symmetry between the parts, not ergodicity within them, produces equal temperatures, equally well for the chaotic stadium and the integrable circle.

A detuning breaks the symmetry and separates $|a,b\rangle$ from $|b,a\rangle$ by an energy $\Delta$ [Fig.~\ref{fig:schematic}(b)]. Each channel then behaves as an off-resonant two-level system: if its coupling $V$ exceeds $\Delta$ it still shares the energy, while for $\Delta\gg V$ the eigenstates remain close to product states and only a small fraction of the imbalance of the channel is shared, $\zeta=1/[1+(\Delta/2V)^2]$ for an isolated channel. Equalization thus requires the cavities to be identical within a tolerance set by the coupling: $\Delta$ grows linearly with the detuning [Eq.~(\ref{eq:A5})], so the channels close one by one as it grows (SM, Sec.~S5). For a detuning of $2\%$ most populated channels have $\Delta/2V\gg1$, which keeps the temperatures apart.

An additive spectrum can also prevent equalization. In a rectangle the motions along $x$ and $y$ are independent and each conserves its own energy, so the energy is their sum, $\varepsilon=f(n_x)+g(n_y)$: exchanging only the $y$ quantum numbers of the two particles turns the channel $\{(n_1,m_1),(n_2,m_2)\}$ into $\{(n_1,m_2),(n_2,m_1)\}$, with exactly the same total energy. The same independence factorizes the boundary functions, so the two channels look identical from the window: the contact couples only to their sum, and their difference is a dark state that stays degenerate with the antisymmetric Bell states, never oscillates, and keeps half of the imbalance of the channel [Fig.~\ref{fig:schematic}(c); Eqs.~(\ref{eq:A10})--(\ref{eq:A13})]. Counting these identities [Eq.~(\ref{eq:A14})] gives exactly the $1100$ to $1600$ dark states of the rectangles, and none for the stadium and the circle. Integrability, or separability, is not the criterion: the circle conserves the angular momentum and separates in polar coordinates, but its radial energy depends on the angular momentum, $\varepsilon=j_{m,n}^2$, so its spectrum is not additive.

The same picture gives the direction of the flow. For two thermal states, with populations $p_i(a)\propto e^{-\varepsilon_a/T_i}$, every channel carries energy from the hot to the cold cavity [Eq.~(\ref{eq:A9})], in line with the known result that heat flows from hot to cold between thermal states unless they start correlated \cite{Partovi2008,Jennings2010}. The net flow is also from hot to cold for pure states with these populations and random phases, and for all the coherent states we studied (SM, Sec.~S6). The direction thus holds whether or not the channels are on resonance; equalization requires the resonance.

\begin{figure}[t]
\includegraphics[width=\columnwidth]{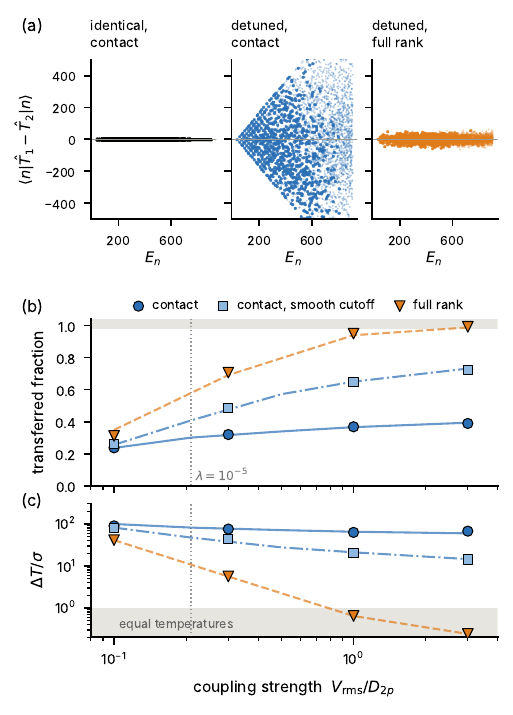}
\caption{\textbf{Few modes, no eigenstate thermalization.} (a) $\langle n|\hat T_1-\hat T_2|n\rangle$, the temperature difference in each eigenstate, against its energy $E_n$ (dark points: eigenstates populated by the initial state). Left: identical stadia with the contact; it vanishes by symmetry. Center: stadia detuned by $2\%$ with the contact; it scatters over a wide range. Right: the same stadia with a random coupling of full rank; it collapses onto a narrow band, as eigenstate thermalization requires. All at $V_{\rm rms}/D_{2p}=3$. (b) Transferred fraction $1-\eta/\eta_0$ and (c) temperature gap in units of the fluctuations, against the coupling strength, for the detuned stadia. Circles: the contact; squares: the contact with a smooth cutoff (SM, Sec.~S9); triangles: the random coupling of full rank. Symbols: time evolution; lines: prediction from the eigenstates. Shaded: equal temperatures within the fluctuations.}
\label{fig:coupling}
\end{figure}

\emph{Few modes, no eigenstate thermalization.}---The late-time temperatures are set by the diagonal elements $\langle n|\hat T_1-\hat T_2|n\rangle$, the temperature difference in each eigenstate, weighted by the initial populations. Eigenstate thermalization (ETH) would make them a smooth function of the energy, the same for all eigenstates of similar energy \cite{Dalessio2016}. Figure~\ref{fig:coupling}(a) shows that this is not what happens [Eqs.~(\ref{eq:A17})--(\ref{eq:A18})]. For identical cavities they vanish for every eigenstate by symmetry: the temperatures equalize although the eigenstates are not thermal. For detuned cavities the eigenstates stay close to the uncoupled product states $|a,b\rangle$, so that $\langle n|\hat T_1-\hat T_2|n\rangle\simeq\varepsilon_a-\varepsilon'_b$ takes widely different values, of either sign, for neighboring eigenstates, and ETH fails. Only a coupling of full rank makes them collapse onto a narrow band.

The reason is that the contact has few modes. It sees each product state $|a,b\rangle$ only through the pressure profile it produces on the window, $\partial_n\phi_a(y)\,\partial_n\phi'_b(y)$, and couples two states through the overlap of their profiles. A window of width $w$ can hold only about $wK/\pi$ independent profiles with wave numbers up to $K$. For the states populated here, $K=\sqrt{2E}$ gives about nine, and the numerical count is $r_{\rm shell}=11$ (SM, Sec.~S9). Thousands of states interact through about ten modes, like the transverse channels of a constriction \cite{Rego1998,Schwab2000}.

Making the contact stronger does not help [Fig.~\ref{fig:coupling}(b),(c)]. We measure its strength by $V_{\rm rms}/D_{2p}$, the typical matrix element over the mean spacing of the two-particle levels (SM, Sec.~S7). Raising it to $3$, where the interaction energy is still below $1\%$ of the total, the transferred fraction saturates below one and the temperature gap stays more than ten times the fluctuations. A stronger contact adds no modes: it freezes the few combinations it acts on, as in a quantum Zeno subspace \cite{Facchi2002}, and the remaining states, which evolve under $H_{\rm eff}=QH_0Q$ with $Q$ the projector out of them [Eq.~(\ref{eq:A16})], exchange only part of the imbalance. Strength widens the tolerance to detuning only up to a point: at the strongest coupling $10^{-4}$ equalizes, but $10^{-3}$ does not (SM, Sec.~S9). More modes help, but slowly: a wider window raises the transfer slightly, a wall that responds nonlocally across the window has fifty of them and transfers barely more (SM, Sec.~S9), and in a model of two random-matrix subsystems the transfer is complete only with about a hundred (SM, Sec.~S8). A coupling that reaches all the states of each energy shell mixes them uniformly \cite{Ponomarev2012}, producing random eigenstates \cite{Tomsovic2018}, ETH and equal temperatures; a window a few wavelengths wide cannot.

\emph{Conclusions.}---Two quantum cavities in contact obey Clausius: in every case we studied, heat flows from hot to cold. What is not guaranteed is that it keeps flowing until the temperatures are equal. An analytical model shows why. Energy crosses the window in quanta, between pairs of levels, and the temperatures equalize only if these pairs are resonant and form Bell states. Exchange symmetry between identical cavities makes them so, exactly and for any coupling. A detuning of one part in a thousand, or the additive spectrum of a rectangle, is enough to prevent it, and a stronger contact does not repair it: a window a few wavelengths wide offers only a few modes, whatever its strength. Where a constriction limits how fast heat flows, such a contact limits how far it gets.

None of the usual signatures of thermalization decides the outcome. The entanglement entropy saturates in all cases; chaotic and integrable cavities equalize or fail alike; and the temperatures equalize without eigenstate thermalization, or fail to despite strong coupling. What decides is the contact: how many modes it has, and whether the levels it connects are resonant.

The mechanism is not specific to billiards: it should hold for any two finite quantum systems coupled through a contact of low rank in their product space, which for one particle per cavity is a contact with few modes, and it does not weaken at higher energy. With many particles the two conditions differ, and whether the bottleneck survives is open (SM, Sec.~S8). Cold atoms offer a natural test: two atomic reservoirs have been joined through a quantum point contact with a tunable number of modes \cite{Krinner2015}, and their temperatures equilibrate when the clouds are large \cite{Brantut2013}; with one atom per reservoir and detuned traps, a single-mode contact should leave the temperatures apart. Two nuclear spins in thermal contact \cite{Micadei2019} are a single channel, the minimal version of the mechanism; coupled microwave billiards realize the window itself, with a single wave \cite{Zhang2026}.

\begin{acknowledgments}
This work was supported by CONICET and CNEA (Argentina). Specifically, the authors acknowledge financial support from CONICET under Grant No. PIP 2022-2024 GI - 11220210100208CO.

\textit{AI use disclosure.}---In accordance with the APS policy on the appropriate use of AI tools, we disclose that the large language model Claude (Anthropic; Claude Opus 5, used through the \texttt{claude.ai} interface and through Claude Code, 2026) assisted this work: in organizing the literature, in sharpening the formulation of the arguments, and in drafting and revising the text of this Letter and of the Supplemental Material; its assistance with the auxiliary numerical scripts and the figures is described in Sec.~S1 of the Supplemental Material. The physical model, the analytical derivation and the numerical methods are the authors' own. The authors directed every step, verified every number, figure and statement against the underlying calculations and the cited sources, and take full responsibility for the content of this Letter.

\textit{Conflict of interest.}---The authors declare that they have no known competing financial interests or personal relationships that could have appeared to influence the work reported in this Letter.
\end{acknowledgments}

\textit{Data availability.}---The data that support the findings of this Letter, and the code that generates them, are not currently publicly available. They will be deposited in a public repository upon publication, and are available from the authors upon reasonable request.

\clearpage
\section*{End Matter}
\setcounter{equation}{0}
\renewcommand{\theequation}{A\arabic{equation}}
\renewcommand{\theHequation}{A\arabic{equation}}

\emph{Appendix A: Channels and the two-level law.}---Without coupling the eigenstates are products $|a,b\rangle$, with energy $E^{(0)}_{ab}=\varepsilon_a+\varepsilon'_b$ ($\varepsilon$ and $\varepsilon'$ are the levels of cavities 1 and 2). Inside a billiard the energy is purely kinetic, so the temperature operators are diagonal in this basis, $\hat T_1|a,b\rangle=\varepsilon_a|a,b\rangle$ and $\hat T_2|a,b\rangle=\varepsilon'_b|a,b\rangle$. Treating a channel $\{|a,b\rangle,|b,a\rangle\}$ as isolated, i.e., neglecting its couplings to other product states (an approximation discussed below),
\begin{equation}
H_{\rm ch}=\begin{pmatrix}E_{ab}&V_{ab}\\ V_{ab}&E_{ba}\end{pmatrix},
\label{eq:A1}
\end{equation}
with $V_{ab}=\lambda\langle a,b|H_{\rm int}|b,a\rangle$, $\Delta_{ab}=E_{ab}-E_{ba}$, and $E_{ab}$ including the diagonal shift $\lambda\langle a,b|H_{\rm int}|a,b\rangle$. Its eigenstates are
\begin{equation}
\begin{split}
|+\rangle&=\cos\theta\,|a,b\rangle+\sin\theta\,|b,a\rangle,\\
|-\rangle&=-\sin\theta\,|a,b\rangle+\cos\theta\,|b,a\rangle,
\end{split}
\label{eq:A2}
\end{equation}
with $\tan2\theta=2V_{ab}/\Delta_{ab}$ and splitting $\Omega=(\Delta_{ab}^2+4V_{ab}^2)^{1/2}$. Starting from $|a,b\rangle$, the energy of cavity 1 performs a Rabi oscillation,
\begin{equation}
E_1(t)=\varepsilon_a-(\varepsilon_a-\varepsilon_b)\,\zeta_{ab}\,\sin^2(\Omega t/2),
\label{eq:A3}
\end{equation}
and, on average over time, the channel removes the fraction
\begin{equation}
\zeta_{ab}=\sin^22\theta=\frac{1}{1+x_{ab}^2},\qquad x_{ab}=\frac{\Delta_{ab}}{2V_{ab}},
\label{eq:A4}
\end{equation}
of its imbalance. A small shape detuning $\delta l=l_{s,2}-l_{s,1}$ shifts each level linearly, $\varepsilon'_i\simeq\varepsilon_i+v_i\,\delta l$ with $v_i=\partial\varepsilon_i/\partial l_s$ (away from avoided crossings), so, to leading order in $\delta l$,
\begin{equation}
\Delta_{ab}\simeq(v_b-v_a)\,\delta l,\qquad x_{ab}\simeq\frac{(v_b-v_a)\,\delta l}{2V_{ab}},
\label{eq:A5}
\end{equation}
so that $|x_{ab}|=|\delta l|/\delta l^*_{ab}$ with $\delta l^*_{ab}=2|V_{ab}|/|v_b-v_a|$ (the difference of the diagonal shifts is of higher order). The sign of $x_{ab}$ matters only in the coherence term of Eq.~(\ref{eq:A6}).
Each channel closes at its own $\delta l^*_{ab}$ (SM, Sec.~S5). For an arbitrary initial state in the channel, with populations $p=|c_{ab}|^2$, $p'=|c_{ba}|^2$ and coherence $\rho=c_{ab}c_{ba}^*$, the diagonal ensemble of $H_{\rm ch}$ gives, to leading order in the detuning,
\begin{equation}
\begin{split}
\langle\hat T_1-\hat T_2\rangle_\infty=(\varepsilon_a-\varepsilon_b)\Big[&\frac{x^2}{1+x^2}(p-p')\\
&+\frac{x}{1+x^2}\,2\,\mathrm{Re}\,\rho\Big],
\end{split}
\label{eq:A6}
\end{equation}
against $(\varepsilon_a-\varepsilon_b)(p-p')$ initially. The isolated-channel approximation is qualitative: the exchange couplings $V_{ab}$ are comparable to the mean two-particle spacing $D_{2p}$, so each channel also mixes with other product states. Summed over the populated channels, Eq.~(\ref{eq:A6}) follows the exact transfer within $0.15$ while their typical $|x|$ is of order one or less, and underestimates it beyond, where other product states contribute (SM, Sec.~S5).

\emph{Appendix B: Identical cavities.}---Let $S|a,b\rangle=|b,a\rangle$. For identical cavities the contact is symmetric, $SH_{\rm int}S=H_{\rm int}$, so $[S,H]=0$ and the eigenstates can be chosen with $S|n\rangle=\pm|n\rangle$. Since $S\hat T_1S=\hat T_2$ (as in Ref.~\cite{Ponomarev2011}),
\begin{equation}
\langle n|\hat T_1|n\rangle=\langle n|S\hat T_1S|n\rangle=\langle n|\hat T_2|n\rangle
\label{eq:A7}
\end{equation}
for every $n$. The same argument gives $\langle n|\hat T_1-\hat T_2|m\rangle=0$ for any two eigenstates of equal parity. Hence, if no symmetric eigenstate is exactly degenerate with an antisymmetric one, the late-time average gives $T_1=T_2$ for any initial state and any symmetric coupling. Exact degeneracies between the two sectors are precisely what the dark states of Appendix D provide. In the channel language $\Delta_{ab}=0$, $\theta=\pi/4$, and both terms of Eq.~(\ref{eq:A6}) vanish, the coherence included. The populations keep the memory of the initial state: for isolated channels $P^\infty_{ab}=P^\infty_{ba}=(p+p')/2$, and for a product state
\begin{equation}
\begin{split}
p_1^\infty(a)&=\sum_b\frac{p_1(a)p_2(b)+p_1(b)p_2(a)}{2}\\
&=\frac{p_1(a)+p_2(a)}{2}=p_2^\infty(a),
\end{split}
\label{eq:A8}
\end{equation}
the arithmetic mean of Ref.~\cite{Ponomarev2011}, not a Gibbs distribution (it holds within a total-variation distance of $0.03$ in the full dynamics). For identical cavities $\chi^{(1)}_y=\chi^{(2)}_y$, so the pressure coupling is a sum of projectors onto symmetric states and $H_{\rm int}|B_-\rangle=0$ for every $|B_-\rangle=(|a,b\rangle-|b,a\rangle)/\sqrt2$: the antisymmetric Bell states are exact eigenstates at the uncoupled energy.

\emph{Appendix C: Direction of the flow.}---Neglecting the coherence in Eq.~(\ref{eq:A6}), whose phase varies from channel to channel, and using that $E_1+E_2$ is conserved within the channel to leading order in the detuning, the energy gained by cavity 1 through the channel $\{a,b\}$ ($\varepsilon_a>\varepsilon_b$) is
\begin{equation}
\delta E_1=-\tfrac12\,\zeta_{ab}\,\big[p_1(a)p_2(b)-p_1(b)p_2(a)\big](\varepsilon_a-\varepsilon_b).
\label{eq:A9}
\end{equation}
Its sign is fixed by the initial populations alone, since $\zeta_{ab}\ge0$. For two Gibbs states with $T_2>T_1$, $p_1(a)p_2(b)-p_1(b)p_2(a)\propto e^{-\varepsilon_a/T_1-\varepsilon_b/T_2}-e^{-\varepsilon_b/T_1-\varepsilon_a/T_2}<0$, so every channel heats the cold cavity. For the coherent states used in the time evolutions, which are not thermal, the net flow is hot to cold in all 78 runs. Initial correlations break the factorization of the populations and with it the argument.

\emph{Appendix D: Dark states in rectangles.}---In a rectangle $L_x\times L_y$, $\varepsilon_{n,m}=\pi^2(n^2/L_x^2+m^2/L_y^2)$, and on a wall $x=$ const the normal derivative factorizes,
\begin{equation}
\partial_n\phi_{n,m}(y)=\pm\frac{2}{\sqrt{L_xL_y}}\,\frac{n\pi}{L_x}\,\sin\frac{m\pi y}{L_y}.
\label{eq:A10}
\end{equation}
For $a=(n_1,m_1)$, $b=(n_2,m_2)$, the channel $c=(n_1,m_2)$, $d=(n_2,m_1)$ has exactly the same energy, $\varepsilon_a+\varepsilon_b=\varepsilon_c+\varepsilon_d$, and, by Eq.~(\ref{eq:A10}), the same boundary function up to a sign $\sigma=\pm1$,
\begin{equation}
\begin{split}
F_{ab}(y)&\equiv\partial_n\phi_a\,\partial_n\phi_b=\sigma F_{cd}(y)\\
&\propto n_1n_2\sin\frac{m_1\pi y}{L_y}\sin\frac{m_2\pi y}{L_y}.
\end{split}
\label{eq:A11}
\end{equation}
In the symmetric sector the contact acts through $M_{kl}=2\lambda\int_wF_kF_l\,dy$, which for the pair is $\mu\,(\begin{smallmatrix}1&\sigma\\\sigma&1\end{smallmatrix})$ with $\mu=2\lambda\int_wF_{ab}^2dy$. Hence
\begin{equation}
|{\rm br}\rangle=\frac{|B^{ab}_+\rangle+\sigma|B^{cd}_+\rangle}{\sqrt2},\qquad |D\rangle=\frac{|B^{ab}_+\rangle-\sigma|B^{cd}_+\rangle}{\sqrt2}.
\label{eq:A12}
\end{equation}
The bright state is shifted by $2\mu$ to first order. The dark state is exact: since $F_{ab}-\sigma F_{cd}=0$ at every point of the window, $H_{\rm int}|D\rangle=0$, and $|D\rangle$ is an eigenstate at the uncoupled energy $E_0$ for any $\lambda$ and any window on that wall. It is degenerate with the antisymmetric state $|B_-^{ab}\rangle$, their coherence never dephases, and since $(\hat T_1-\hat T_2)|B_-^{ab}\rangle=(\varepsilon_a-\varepsilon_b)|B_+^{ab}\rangle$, a channel started in $|a,b\rangle$ keeps
\begin{equation}
\langle\hat T_1-\hat T_2\rangle_\infty=f\,(\varepsilon_a-\varepsilon_b),
\label{eq:A13}
\end{equation}
with $f$ the weight of $|B_+^{ab}\rangle$ in the dark subspace; for a pair $f=1/2$, and the channel shares half of its imbalance [Fig.~\ref{fig:schematic}(c)]. In general, the dark states of a group $g$ of $m_g$ degenerate symmetric states are the combinations whose boundary function vanishes on the window, so
\begin{equation}
N^{(g)}_{\rm dark}=m_g-\mathrm{rank}\{F_k(y)\}_{k\in g},
\label{eq:A14}
\end{equation}
which reproduces exactly the numbers of dark states found numerically: 1146, 1110, 1619, 1584 and 1557 for the $\pi$, golden-ratio, $\sqrt2$, $2:1$ and square rectangles, none for the stadium and the circle. The circle, integrable but with a nonadditive spectrum of Bessel zeros, has no degenerate groups.

\emph{Appendix E: Strong coupling.}---The contact is a positive sum of projectors, $H_{\rm int}=\int_wdy\,|\chi^{(1)}_y\chi^{(2)}_y\rangle\langle\chi^{(1)}_y\chi^{(2)}_y|$, with $\langle i|\chi_y\rangle=\partial_n\phi_i(y)$; numerically
\begin{equation}
H_{\rm int}=\sum_{j=1}^{r}s_j\,|g_j\rangle\langle g_j|,\qquad r=17,
\label{eq:A15}
\end{equation}
the next eigenvalue being at least eight orders of magnitude smaller. Strictly, the discretized operator has a larger rank, but the remaining directions are negligible as long as $\lambda s_j\ll D_{2p}$ for $j>r$, which holds for all couplings studied here and up to $\lambda\sim10^{8}$ times larger; $r$ is thus an effective rank. What the argument below uses is this rank in the product space, not the number of modes of the contact: for one particle per cavity the pressure at a point is a projector, $\hat P(y)=|\chi_y\rangle\langle\chi_y|$, so a coupling with few modes has low rank, while for many particles per cavity it need not (SM, Sec.~S8). The rank grows with $N_1$, because the strongest directions come from the highest levels kept, while the rank within the energy shell of the initial state, $r_{\rm shell}=11$, does not (SM, Sec.~S9). For a positive perturbation of rank $r$, Weyl's interlacing theorem bounds the eigenvalues (here up to the negligible remainder), $E^{(0)}_k\le E_k\le E^{(0)}_{k+r}$, for any $\lambda$: the spectrum is only rearranged locally. When $\lambda s_j$ exceeds the energy range of the initial state for $j\le r$, while it stays negligible for $j>r$, $r$ eigenstates follow the $|g_j\rangle$ to high energy, and the others approach the eigenstates of
\begin{equation}
H_{\rm eff}=QH_0Q,\qquad Q=1-\sum_{j=1}^{r}|g_j\rangle\langle g_j|,
\label{eq:A16}
\end{equation}
which does not depend on $\lambda$ in this effective strong-coupling regime. The transfer therefore approaches a plateau: for $N_1=80$, Eq.~(\ref{eq:A16}) gives an upper limit of $0.43$ for $l_{s,2}=0.51$ and $0.98$ for $0.50005$, which the finite-coupling values approach ($0.39$ and $0.97$ at $V_{\rm rms}/D_{2p}=3$; Fig.~\ref{fig:coupling}); with the smooth cutoff, converged in $N_1$, the limits are $0.76$ and $0.98$. A random coupling of full rank has $Q=0$: no subspace is protected, the eigenstates at strong coupling are random in the product basis, $\langle n|\hat T_1-\hat T_2|n\rangle\to0$ as eigenstate thermalization requires, and the temperatures equalize. The value of this limit depends on the cutoff of the contact, but stays below one, with $\Delta T/\sigma\gg1$, for every cutoff we tested (SM, Sec.~S9).

\emph{Appendix F: Diagonal ensemble and eigenstate thermalization.}---Without exact degeneracies the late-time temperatures are
\begin{equation}
T_1-T_2\big|_\infty=\sum_n|c_n|^2\,\langle n|\hat T_1-\hat T_2|n\rangle ,\qquad c_n=\langle n|\psi_0\rangle,
\label{eq:A17}
\end{equation}
which depends on the initial state through $|c_n|^2$. ETH states that $\langle n|\hat T_1-\hat T_2|n\rangle\simeq f(E_n)$ with fluctuations that vanish with the number of states mixed, so that Eq.~(\ref{eq:A17}) depends only on the energy. In the isolated-channel picture, Eq.~(\ref{eq:A2}) gives, to leading order in the detuning,
\begin{equation}
\langle\pm|\hat T_1-\hat T_2|\pm\rangle=\pm(\varepsilon_a-\varepsilon_b)\cos2\theta=\pm\frac{(\varepsilon_a-\varepsilon_b)\,x}{\sqrt{1+x^2}} .
\label{eq:A18}
\end{equation}
For $x=0$ (identical cavities) the elements vanish exactly, without any smoothness in energy: equality by symmetry, not by ETH. For $x\gg1$ (detuned) they take the values $\varepsilon_a-\varepsilon'_b$, of either sign, which change by hundreds from one eigenstate to the next [Fig.~\ref{fig:coupling}(a)]: ETH fails. A coupling that mixes $\mathcal N$ product states with random amplitudes gives elements of zero mean and fluctuations of order $(\varepsilon_a-\varepsilon_b)/\sqrt{\mathcal N}$, as for random states, restoring ETH; a contact of rank $r$ cannot do so, since by Eq.~(\ref{eq:A16}) the eigenstates outside its $r$ directions are those of $QH_0Q$, close to products.

\clearpage
\setcounter{figure}{0}\renewcommand{\thefigure}{S\arabic{figure}}
\setcounter{table}{0}\renewcommand{\thetable}{S\arabic{table}}
\setcounter{equation}{0}\renewcommand{\theequation}{S\arabic{equation}}
\renewcommand{\theHequation}{S\arabic{equation}}
\renewcommand{\theHfigure}{S\arabic{figure}}
\renewcommand{\theHtable}{S\arabic{table}}

\section*{Supplemental Material}

The analytical theory of the mechanism (channels, the symmetry argument, the direction of the flow, dark states and the strong-coupling limit) is given in the End Matter of the main text; this Supplemental Material contains the numerical details and verifications. The sections follow the order of the main text: the numerical model (S1) and the coupling (S2); the temperature and the criterion for equal temperatures (S3); entanglement and level statistics (S4); the mechanism, including the direction of the heat flow (S5); other initial states (S6); the role of the coupling strength and of the number of modes (S7); a generic random-matrix model (S8); and the ultraviolet cutoff of the contact (S9).

\section*{S1. NUMERICAL MODEL}
\label{sec:S1}

The eigenstates \(\phi_i\) and energies \(\varepsilon_i\) of each cavity are computed with the scaling method \cite{Vergini1995}; for the circle (a quarter circle, as in the main text) and the rectangles they reproduce the analytic levels to \(10^{-13}\) or better. We use \(\hbar=1\), \(m=1/2\), cavities of unit linear size, and keep \(N_1=80\) levels per cavity (dimension \(6400\)); runs with \(N_1=120\) are used to check finite-size effects. The window occupies \(0.1\le y\le0.8\) of the common wall. The normal derivatives on the window are analytic for the rectangles and are obtained from the first interior grid point for the stadium and the circle, with an error of about \(1\%\) for the highest levels.

The initial state is a product of Gaussian coherent states (width \(\omega=200\)), centered at \(\mathbf Q_1=(-0.2,0.5)\) with momentum \(\mathbf P_1=(-4,4)\) in cavity 1 and at \(\mathbf Q_2=(0.3,0.5)\) with \(\mathbf P_2=(8,-8)\) in cavity 2. The coherent states are projected onto the \(N_1=80\) levels of each cavity and renormalized; the weight left outside the basis is below \(1\%\) except in the elongated rectangles, whose larger area places more levels below the energy of the initial state. The total Hamiltonian is diagonalized exactly and the state is evolved in its eigenbasis up to \(t/\tau\sim10^6\), with \(\tau\) the mean collision time; late-time averages are taken over \(t/\tau>10^5\).

\paragraph{Late-time predictions.} The residual imbalance is \(\eta=|\langle T_1\rangle-\langle T_2\rangle|/(\langle T_1\rangle+\langle T_2\rangle)\). If the relative phases of all populated eigenstates randomize, its late-time value is given by the diagonal ensemble,
\begin{equation}
\eta_{\rm diag}=\frac{\bigl|\sum_n |c_n|^2\langle n|\hat T_1-\hat T_2|n\rangle\bigr|}{\sum_n |c_n|^2\langle n|\hat T_1+\hat T_2|n\rangle},\qquad c_n=\langle n|\psi_0\rangle .
\label{eq:etadiag}
\end{equation}
When the spectrum has exactly degenerate groups of eigenstates, their relative phases never randomize and the coherences inside each group must be kept:
\begin{equation}
\eta_{\rm gen}=\frac{\bigl|\sum_g \langle\psi_0|P_g(\hat T_1-\hat T_2)P_g|\psi_0\rangle\bigr|}{\sum_n |c_n|^2\langle n|\hat T_1+\hat T_2|n\rangle},
\label{eq:etagen}
\end{equation}
with \(P_g\) the projector onto each degenerate group (energy difference below \(10^{-8}\)). Unlike \(\eta_{\rm diag}\), \(\eta_{\rm gen}\) does not depend on the arbitrary basis that the numerical diagonalization chooses inside a degenerate group. Throughout, the ``fraction of the imbalance transferred'' is \(1-\eta/\eta_0\), with \(\eta_0\) the initial imbalance.

\paragraph{Computational tools and AI assistance.} The eigenstates of the individual cavities are obtained with the authors' own implementation of the scaling method, and the coupling matrix and the time evolutions with the authors' own programs; the diagonalizations use standard LAPACK routines. The auxiliary scripts written for the present work -- the convergence and robustness scans of Secs.~S3 and S9, a Python reimplementation of the time evolution and of the diagonal ensembles of Eqs.~(\ref{eq:etadiag}) and (\ref{eq:etagen}), used to cross-check the authors' original Fortran codes, and the figures -- were written with the assistance of the large language model Claude (Anthropic; Claude Opus 5, used through the \texttt{claude.ai} interface and through Claude Code, 2026), under the direction of the authors. The results were checked against independent calculations and analytic limits wherever available: the levels of the circle and the rectangles against their closed forms, the two-level law of Eq.~(A6) of the End Matter against the exact gap, the late-time averages against the diagonal ensembles, and every figure against the raw output files. The authors verified all of it and take full responsibility for the results.

\section*{S2. THE COUPLING}
\label{sec:S2}

\paragraph{Pressure on the wall.} Since the wave functions vanish on the walls, the particles can only interact through the normal derivatives of their wave functions. The force per unit length that a particle exerts on the wall at the point \(y\) is \(P(y)=|\partial_n\psi(y)|^2\) (in our units), as follows from the Hellmann--Feynman theorem for a local displacement of the wall \cite{Barnett2000}. The corresponding operator has matrix elements \(\langle i|\hat P(y)|k\rangle=\partial_n\phi_i(y)\,\partial_n\phi_k(y)\), and the coupling of the main text is
\begin{equation}
\begin{split}
H_{\rm int}&=\int_w dy\;\hat P_1(y)\otimes\hat P_2(y),\\
\langle i,j|H_{\rm int}|k,l\rangle&=\int_w dy\,\partial_n\phi^{(1)}_i\partial_n\phi^{(1)}_k\,\partial_n\phi^{(2)}_j\partial_n\phi^{(2)}_l .
\end{split}
\label{eq:hintPSM}
\end{equation}

\paragraph{Where the product of pressures comes from.} A simple picture motivates this form. Imagine that the part of the wall occupied by the window can move: let \(u(y)\) be its small displacement, positive towards cavity 2, with an elastic restoring energy \((\kappa/2)\int u^2dy\). When the wall moves by \(u\), cavity 1 becomes slightly larger and the energy of its particle decreases by \(\int P_1u\,dy\); cavity 2 becomes smaller and the energy of its particle increases by \(\int P_2u\,dy\). To first order in \(u\),
\begin{equation}
H=H_1+H_2+\int_w dy\,\Bigl[\frac{\kappa}{2}u^2-u\,(\hat P_1-\hat P_2)\Bigr].
\label{eq:membrane}
\end{equation}
If the wall is stiff and fast compared with the particles, it follows them adiabatically and sits at the minimum of Eq.~(\ref{eq:membrane}) with respect to \(u\): \(\kappa u=\hat P_1-\hat P_2\), i.e., the wall is displaced in proportion to the pressure difference across it. Substituting this value back,
\begin{equation}
\begin{split}
H=&\Bigl[H_1-\frac{1}{2\kappa}\int_w\hat P_1^2\Bigr]+\Bigl[H_2-\frac{1}{2\kappa}\int_w\hat P_2^2\Bigr]\\
&+\frac{1}{\kappa}\int_w dy\,\hat P_1\hat P_2 .
\end{split}
\label{eq:eliminated}
\end{equation}
The first two brackets involve one cavity each: they describe how a movable wall modifies each cavity separately, and they exchange no energy between them. The last term is the only one that acts on both particles; it is Eq.~(\ref{eq:hintPSM}) with \(\lambda=1/\kappa\), and it is the term responsible for the energy transfer. We keep only this term. The single-cavity terms are not small in this static picture, and they grow with the number of levels kept, because the static elimination of \(u\) is not valid for fast modes; a full treatment would include the dynamics of the wall. Being single-particle operators, they can be absorbed in \(H_1\) and \(H_2\), which we take to describe each cavity with its actual wall; the detuning is then defined on these physical levels. Including them explicitly leaves the exchange symmetry exact: for identical cavities \(\Delta_{ab}\) stays zero to \(10^{-14}\) and the transfer is complete. We therefore regard Eq.~(\ref{eq:eliminated}) as a motivation of the form of the coupling, not as a derivation of the model: Eq.~(\ref{eq:hintPSM}) is an effective model of a local contact. What our conclusions rely on is its structure, a sum of a few projectors on boundary functions, which the other couplings below allow us to test.

\paragraph{A contact has few modes.} Since \(\langle i|\hat P(y)|k\rangle\) is the product of a function of \(i\) and the same function of \(k\), \(\hat P(y)=|\chi_y\rangle\langle\chi_y|\) with \(\langle i|\chi_y\rangle=\partial_n\phi_i(y)\), and
\begin{equation}
H_{\rm int}=\int_w dy\,|\chi^{(1)}_y\rangle|\chi^{(2)}_y\rangle\langle\chi^{(1)}_y|\langle\chi^{(2)}_y| .
\end{equation}
Two consequences follow. First, the coupling acts only on a few independent combinations of the boundary functions on the window: \(99.99\%\) of its weight (the trace, i.e., the sum of its eigenvalues) is carried by 9--17 singular directions, depending on the geometry, out of 6400 (counting instead every eigenvalue above \(10^{-8}\) of the largest gives \(r=17\) for the stadium, the value used in the End Matter, Eq.~(A15)). Restricted to the energy shell of the initial state, \(|E_{ab}-E_0|<2\sigma_E\), the rank is \(r_{\rm shell}=10\)--\(12\) for all geometries and does not depend on \(N_1\) (Sec.~S9). This is the low number of modes discussed in the main text. Second, for identical cavities \(\chi^{(1)}_y=\chi^{(2)}_y\), so each term projects onto a state that is symmetric under the exchange of the particles: \(H_{\rm int}\) commutes with the exchange and annihilates every antisymmetric state, and the \(N_1(N_1-1)/2=3160\) states \((|a,b\rangle-|b,a\rangle)/\sqrt2\) are exact eigenstates of \(H\) (verified to machine precision). This is the analogue of the fact that a contact interaction does not act on spatially antisymmetric two-particle states.

\paragraph{No classical counterpart.} The product of the local pressures has no classical analogue for point particles: it acts only when both particles hit the same point of the window at the same instant, which never happens. A rigid piston, which couples the total pressures on the window, is the \(\ell\to\infty\) limit of the nonlocal wall of Sec.~S9, and it does not equalize the detuned stadia either (transfer \(0.79\) at \(V_{\rm rms}/D_{2p}=3\)). The classical adiabatic piston equilibrates through the fluctuations of a wall with its own dynamics, which the static elimination used here removes.

\paragraph{Other couplings.} To test whether the results depend on the form of the interaction we used four other couplings, not derived from a microscopic picture (we call the contact of Eq.~(\ref{eq:hintPSM}) \(H^{(1)}\equiv H_{\rm int}\)):
\begin{align}
\langle i_1,i_2|H^{(2)}|j_1,j_2\rangle&=(\varepsilon^{(1)}_{i_1}-\varepsilon^{(2)}_{j_2})(\varepsilon^{(2)}_{i_2}-\varepsilon^{(1)}_{j_1}),\label{eq:H3}\\
\langle i_1,i_2|H^{(3)}|j_1,j_2\rangle&=(P^{(1)}_{i_1}-P^{(2)}_{j_2})(P^{(2)}_{i_2}-P^{(1)}_{j_1}),\label{eq:H4}\\
\langle i_1,i_2|H^{(4)}|j_1,j_2\rangle&=\cos(k^{(1)}_{i_1}k^{(2)}_{i_2}k^{(1)}_{j_1}k^{(2)}_{j_2}),\label{eq:H5}
\end{align}
and a real symmetric random matrix \(H^{\rm (rand)}\) of full rank. Here \(P_i=\int_w|\partial_n\phi_i|^2dy\) is the window pressure of level \(i\) and \(k_i=\sqrt{\varepsilon_i}\). Expanding the products shows that \(H^{(2)}\) and \(H^{(3)}\) are sums of four terms of the form (function of the initial state)\(\times\)(function of the final state), so they act on only four directions, like a contact. \(H^{(4)}\) behaves as a pseudo-random matrix that connects thousands of directions, and \(H^{\rm (rand)}\) connects all of them. \(H^{(2)}\), \(H^{(3)}\) and \(H^{(4)}\) respect the exchange symmetry for identical cavities; the random matrix does not. For identical stadia all symmetric couplings give \(\eta_{\rm diag}\lesssim10^{-12}\), and the square keeps \(\eta\simeq0.1\)--\(0.2\) with all of them. The strength of each coupling is measured by \(V_{\rm rms}/D_{2p}\), the root-mean-square matrix element between product states over the mean spacing \(D_{2p}\simeq0.39\) of the two-particle levels in the energy shell of the initial state; \(\lambda=10^{-5}\) corresponds to \(V_{\rm rms}/D_{2p}\simeq0.21\) for the contact. The comparison of the couplings for detuned cavities is given in Sec.~S7.

\section*{S3. TEMPERATURE AND THE CRITERION FOR EQUAL TEMPERATURES}
\label{sec:S3}

\begin{figure}[t]
\includegraphics[width=\columnwidth]{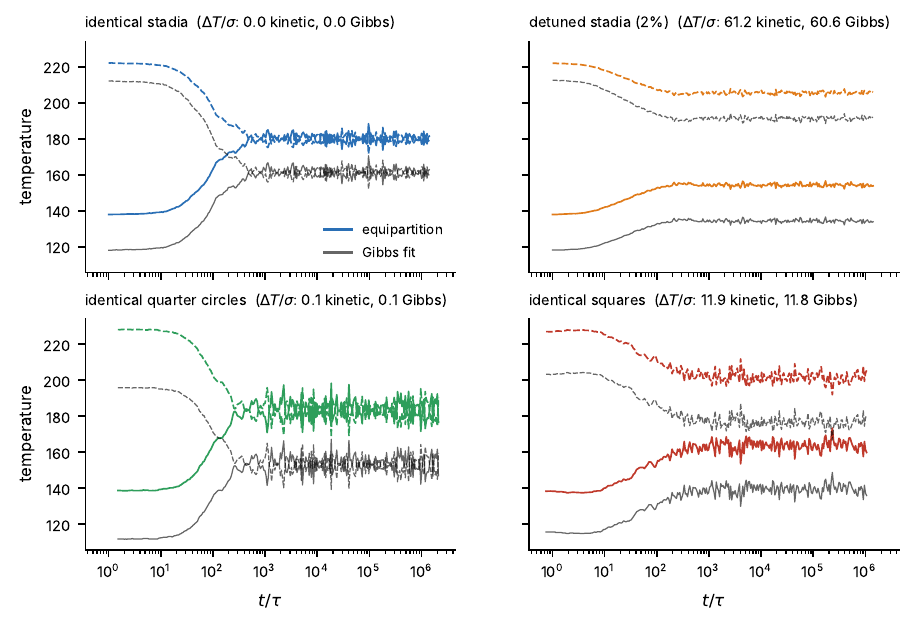}
\caption{\textbf{Two definitions of temperature.} Equipartition temperature (color; solid: cold cavity, dashed: hot cavity) and temperature from a Gibbs fit of the populations (gray), for the four cases of Fig.~1 of the main text (\(\lambda=10^{-5}\), the same runs). The ratio \(\Delta T/\sigma\) obtained with each definition is given in each title.}
\label{fig:S_gibbs}
\end{figure}

\paragraph{Temperature without a Gibbs state.} Following Ref.~\cite{Rivas2025}, the temperature of each cavity is defined by equipartition, \(T_i=\mathrm{Tr}[\rho_i\hat T_i]\), with the temperature operator \(\hat T_i\) equal to the kinetic energy of cavity \(i\) (\(k_B=1\)), which applies in or out of equilibrium. The Gibbs distribution is the particular form taken by the populations of a system in contact with a bath much larger than itself, whose density of states can be expanded as \(\ln\rho_B(E-\varepsilon)\simeq\ln\rho_B(E)-\beta\varepsilon\). For two cavities of the same size neither is a bath for the other: at total energy \(E\) the populations of one cavity follow the density of states of the other, \(p(\varepsilon)\propto\rho_2(E-\varepsilon)\), which for a two-dimensional billiard is nearly constant (Weyl's law), not exponential. A temperature can nevertheless be obtained by fitting the populations of each cavity to a Gibbs form. Figure~\ref{fig:S_gibbs} compares both definitions. The Gibbs temperatures are lower by \(6\)--\(16\%\), but they follow the same history, and where the gap is visible \(\Delta T/\sigma\) agrees within \(3\%\) between the two definitions: the classification of the cases does not depend on how the temperature is defined.

\paragraph{Equal temperatures, but memory of the initial state.} Equal temperatures in identical cavities do not mean that each cavity forgets its initial state. For identical stadia, the late-time populations of each cavity coincide with the average of the two initial distributions, \((|a_i|^2+|b_i|^2)/2\), within a total-variation distance of \(0.03\), for \(\lambda\) from \(10^{-5}\) to \(10^{-1}\). This is the arithmetic-mean regime of Ponomarev \emph{et al.}~\cite{Ponomarev2011}, and it is what the Bell-pair picture predicts: each pair \(|a,b\rangle\), \(|b,a\rangle\) ends up with equal weights.

\begin{figure}[t]
\includegraphics[width=\columnwidth]{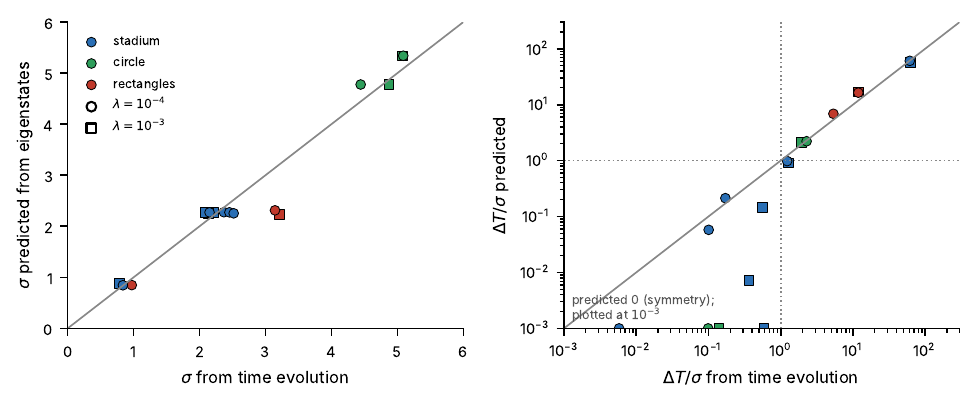}
\caption{\textbf{Predicting the fluctuations.} Left: amplitude \(\sigma\) of the temperature fluctuations predicted from the eigenstates, Eq.~(\ref{eq:sigma}), against the time evolution. Right: \(\Delta T/\sigma\), predicted against measured; dotted lines: \(\Delta T/\sigma=1\). Rectangles: generalized diagonal ensemble. Cases whose prediction vanishes by symmetry are plotted at \(10^{-3}\); for those below one the measured gap is the residue of the finite average, not the prediction.}
\label{fig:S_sigma}
\end{figure}

\paragraph{The criterion.} In a finite isolated system the temperatures keep fluctuating. We compare the late-time gap \(\Delta T=|\langle T_1\rangle-\langle T_2\rangle|\) with the amplitude \(\sigma\) of the fluctuations, \(\sigma^2=(\sigma_{T_1}^2+\sigma_{T_2}^2)/2\), and say that the temperatures are equal when \(\Delta T/\sigma<1\). The two quantities are related by \(\Delta T/\sigma=\eta/[\sigma/(T_1+T_2)]\): \(\eta\) measures the imbalance and is what the mechanism predicts, while \(\Delta T/\sigma\) decides whether it is visible. Both can be predicted from the eigenstates. \(\Delta T\) follows from Eq.~(\ref{eq:etadiag}). For the fluctuations, the temporal variance of \(\hat T_1-\hat T_2\) around its diagonal value is, in the absence of degenerate gaps,
\begin{equation}
\mathrm{Var}_t(\hat T_1-\hat T_2)=\sum_{n\neq m}|c_n|^2|c_m|^2\,|\langle n|\hat T_1-\hat T_2|m\rangle|^2 ,
\label{eq:sigma}
\end{equation}
and since \(T_1-T_2\) fluctuates while \(T_1+T_2\) is nearly constant, \(\sigma\simeq\frac12\sqrt{\mathrm{Var}_t}\). Figure~\ref{fig:S_sigma} compares these predictions with the time evolutions. The predicted \(\sigma\) agrees within \(10\%\) for the stadium and the circle; for the rectangles, whose additive spectrum has degenerate eigenvalues, the coefficients \(c_n\) are not defined basis by basis, and we use the diagonal ensemble generalized to one vector per eigenspace, the normalized projection of the initial state: the predicted \(\sigma\) is then \(1\)--\(10\%\) low instead of \(10\)--\(19\%\), and the predicted \(\Delta T/\sigma\) of the identical squares goes from \(1.1\)--\(1.7\) to \(12.9\), against \(11.8\) measured. Over the cases with \(\Delta T/\sigma\ge1\) the ratio of predicted to measured has median \(1.0\) and lies between \(0.72\) and \(1.13\); the predicted \(\Delta T/\sigma\) classifies all cases as the time evolution does, except \(l_{s,2}=0.50005\), which lies at the border (\(0.9\)--\(1.0\) predicted, \(1.2\)--\(1.3\) in the dynamics). Where the prediction is zero by symmetry, the time evolution leaves a residual gap compatible with the error of the late-time average over \(N_{\rm eff}\simeq47\) independent samples, \(\sigma/\sqrt{N_{\rm eff}}\simeq0.3\). The fluctuations are themselves a fingerprint of the energy exchange: they are large when energy goes back and forth between the cavities (\(\sigma\simeq2.3\) for identical stadia) and small when it does not (\(\sigma\simeq0.85\) for \(l_{s,2}=0.51\)).

\paragraph{Thermal reference.} For two parts of the same size the thermal prediction is the microcanonical ensemble of the composite system with the energy distribution of the initial state. Giving every product state the initial weight averaged over narrow bins of the two-particle spectrum yields a residual imbalance \(\eta_{\rm mc}\le10^{-3}\) for all the pairs of Fig.~1 of the main text, that is, equal temperatures. Failing to equalize therefore means failing to reach the microcanonical state.

\begin{figure}[t]
\includegraphics[width=0.45\columnwidth]{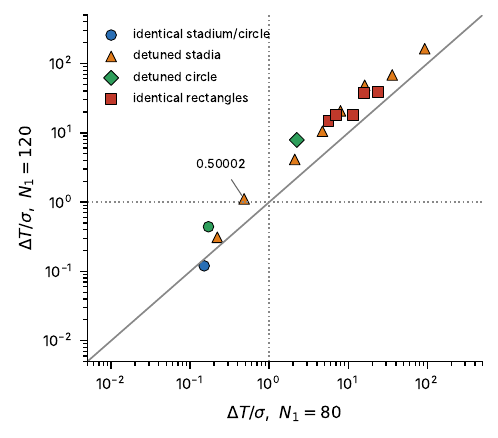}
\caption{\textbf{Finite size.} \(\Delta T/\sigma\) with \(N_1=120\) against \(N_1=80\) levels per cavity (\(\lambda=10^{-5}\)). Dotted lines: \(\Delta T/\sigma=1\).}
\label{fig:S_size}
\end{figure}

\paragraph{Finite size.} Figure~\ref{fig:S_size} compares \(\Delta T/\sigma\) for \(N_1=80\) and \(N_1=120\). The classification is the same for all cases. The transferred fraction of detuned cavities decreases with \(N_1\) at fixed \(\lambda\), which is a property of the cutoff and not of the size: the basis acts as the ultraviolet cutoff of the contact, and refining it weakens the effective coupling within the energy shell (Sec.~S9). With the smooth cutoff the transfer is the same for \(N_1=80\), \(120\) and \(160\). The criterion itself becomes more demanding as the system grows, because \(\sigma\) decreases.

\section*{S4. ENTANGLEMENT AND LEVEL STATISTICS}
\label{sec:S4}

\paragraph{Entanglement entropy.} \(S_{\rm ent}=-\mathrm{Tr}\rho_1\ln\rho_1\) saturates at \(t/\tau\sim10^3\) (Fig.~1 of the main text). The appropriate reference is not the Page value \cite{Page1993} for a random state of the full \(80\times80\) space (\(\approx3.9\)) but that of a random state in the energy shell of the initial state, \(\approx3.1\) for the stadium. The late-time values at \(\lambda=10^{-5}\) are \(2.47\) (identical stadia), \(2.81\) (\(l_{s,2}=0.51\)), \(1.95\) (identical circles) and \(2.08\) (squares), and change by less than \(0.03\) up to \(\lambda=10^{-3}\). A state with the initial moduli \(|\alpha_a||\beta_b|\) and random phases has \(S_{\rm ent}=2.71\) for the stadia, \(2.03\) for the circles and \(2.31\) for the squares. The detuned stadia, whose eigenstates remain close to products, saturate near this value: their entanglement comes from the dephasing of the product components by the nonadditive diagonal shifts \(\lambda\langle a,b|H_{\rm int}|a,b\rangle\), which entangles the cavities without exchanging energy.

\begin{figure}[t]
\includegraphics[width=\columnwidth]{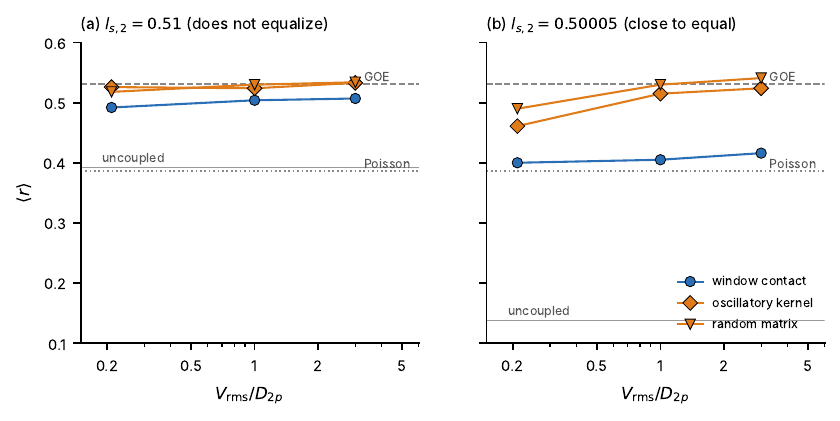}
\caption{\textbf{Level statistics do not decide.} Mean ratio of consecutive spacings \(\langle r\rangle\) of the two-particle levels against the coupling strength, for the window contact and two couplings with many modes. Dashed: GOE; dotted: Poisson; gray: uncoupled cavities.}
\label{fig:S_levels}
\end{figure}

\paragraph{Level statistics.} One might expect the statistics of the two-particle levels to tell which cases equalize. Figure~\ref{fig:S_levels} shows the mean ratio of consecutive level spacings \(\langle r\rangle\) \cite{Oganesyan2007,Atas2013}, which is \(0.386\) for Poisson and \(0.531\) for random-matrix (GOE) statistics, in the central \(60\%\) of the spectrum. Without coupling the two-particle spectrum is a sum of two independent spectra and is close to Poisson, or even more clustered for nearly identical cavities. With the contact, the stadia detuned by \(2\%\) reach \(\langle r\rangle\simeq0.50\), close to GOE, and do not equalize; the stadia detuned by \(10^{-4}\) stay close to Poisson and come close to equal temperatures. Level repulsion measures whether states neighboring in energy are mixed, whereas equalization requires mixing \(|a,b\rangle\) with \(|b,a\rangle\), which are separated by the detuning. The couplings with many modes reach GOE statistics at \(V_{\rm rms}/D_{2p}\gtrsim1\) in both cases. We therefore do not claim that the coupled system is chaotic in the sense of level statistics.

\section*{S5. THE MECHANISM: BELL PAIRS, DARK STATES, AND THE DIRECTION OF THE FLOW}
\label{sec:S5}

\paragraph{Counting Bell pairs and dark states.} An eigenstate is counted as a Bell pair when it is dominated by two product states \(|a,b\rangle\), \(|b,a\rangle\) with nearly equal weights (inverse participation ratio above \(0.45\) and weight ratio above \(0.9\)), and a channel \(\{a,b\}\) is counted as a Bell channel when \(|a,b\rangle\) or \(|b,a\rangle\) dominates a Bell pair. For identical cavities without exact degeneracies every eigenstate is symmetric or antisymmetric, and all channels share their energy equally. In the rectangles, a degenerate group of channels with the same total energy produces combinations of Bell\(^+\) states that the contact does not see: we count them as symmetric eigenstates that remain exactly at an unperturbed energy (dark states). Their origin is the exact identity of the boundary functions of channels related by exchanging the \(y\) quantum numbers, and their number is given by the counting rule of the End Matter, Eq.~(A14), which we evaluated group by group (Table~\ref{tab:groups}); it agrees exactly with the numerical count. There are none in the stadium and the circle, \(1146\) and \(1110\) in the \(\pi\) and golden-ratio rectangles, and \(1557\)--\(1619\) in the square and the \(\sqrt2\) and \(2:1\) rectangles; the numbers do not depend on \(\lambda\). Rational aspect ratios produce large groups (up to 58 channels in the square) and more dark states. Keeping only the coherences inside the degenerate groups, Eq.~(\ref{eq:etagen}), reproduces the time-evolved imbalance of the five rectangles within \(4\%\).

\begin{table*}[t]
\centering\footnotesize
\begin{tabular}{lccccc}
\hline\hline
 & groups of 2 & groups of 3--6 & largest group & \(N_{\rm dark}\) (rule) & \(N_{\rm dark}\) (numerical)\\
\hline
Rect.\ \(\pi\) & 773 & 143 & 6 & 1146 & 1146 \\
Rect.\ golden & 835 & 149 & 5 & 1110 & 1110 \\
Rect.\ \(\sqrt2\) & 32 & 60 & 42 & 1619 & 1619 \\
Rect.\ 2:1 & 20 & 85 & 42 & 1584 & 1584 \\
Square & 12 & 46 & 58 & 1557 & 1557 \\
Stadium, circle & 0 & 0 & -- & 0 & 0 \\
\hline\hline
\end{tabular}
\caption{Degenerate groups of symmetric two-particle states with the same total energy, and dark states predicted by Eq.~(A14) of the End Matter, \(N_{\rm dark}=\sum_g(m_g-\mathrm{rank}\{F_k\}_{k\in g})\), compared with the number of symmetric eigenstates that stay at an unperturbed energy (\(\lambda=10^{-5}\)). The circle has a single accidental pair, with no dark state.}
\label{tab:groups}
\end{table*}

\begin{figure}[t]
\includegraphics[width=\columnwidth]{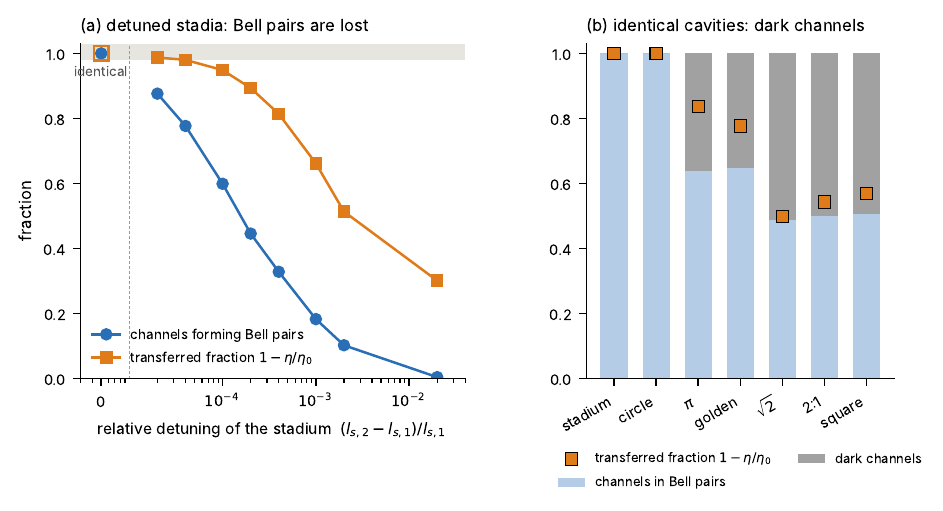}
\caption{\textbf{Bell pairs and dark states control the equalization} (\(\lambda=10^{-5}\)). (a) Detuned stadia: fraction of channels that form Bell pairs (blue) and fraction of the imbalance transferred (orange), against the relative detuning; the identical case is shown at the left. Shaded: equal temperatures within the fluctuations. (b) Identical cavities: fraction of channels in Bell pairs (light) and dark (gray), and fraction of the imbalance transferred (squares).}
\label{fig:S_bell}
\end{figure}

Figure~\ref{fig:S_bell} shows how the Bell channels control the energy transfer. As the detuning of the stadium grows, the fraction of Bell channels falls from \(0.88\) (\(l_{s,2}=0.50001\)) to \(0.004\) (\(0.51\)), and the transferred fraction falls with it; the temperatures are equal within the fluctuations only while most channels remain Bell pairs. In identical cavities, the dark channels of the rectangles (\(35\)--\(50\%\)) leave \(16\)--\(50\%\) of the imbalance untransferred. The transferred fraction always lies above the fraction of Bell channels: channels that are not Bell pairs still share part of their energy, as an off-resonant two-level system does, and states close in energy mix partially. In the few-channel contact considered here, Bell pairs are what complete the equalization; without them, part of the imbalance stays.

\begin{figure}[t]
\includegraphics[width=\columnwidth]{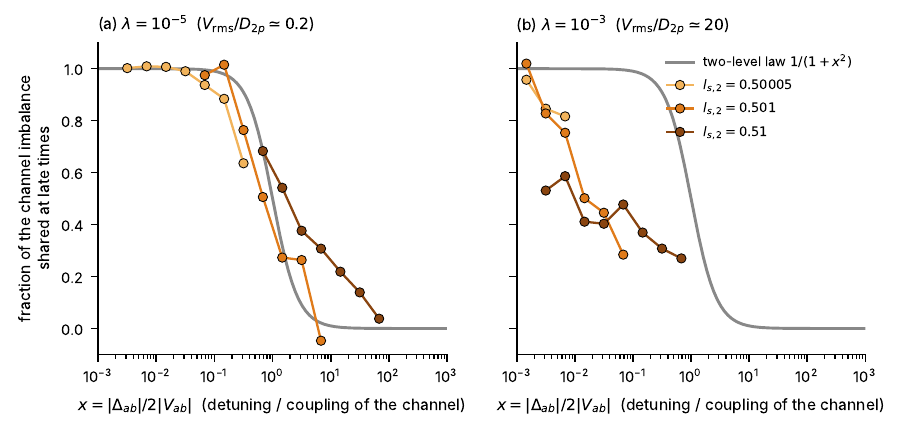}
\caption{\textbf{Energy transferred by each channel} in detuned stadia, averaged over channels with similar ratio \(x\) between detuning splitting and coupling, compared with the two-level law \(1/(1+x^2)\) (gray). Left: \(\lambda=10^{-5}\); right: \(\lambda=10^{-3}\).}
\label{fig:S_channels}
\end{figure}

\paragraph{The two-level law, channel by channel.} Figure~\ref{fig:S_channels} tests the two-state picture of Fig.~2 of the main text pair by pair. For each channel \(\{a,b\}\) we compute the fraction of its initial imbalance that is transferred at late times, and average it over channels with similar \(x=|\Delta_{ab}|/2|V_{ab}|\), where \(\Delta_{ab}\) is the energy difference between \(|a,b\rangle\) and \(|b,a\rangle\) and \(V_{ab}=\lambda\langle a,b|H_{\rm int}|b,a\rangle\). At \(\lambda=10^{-5}\) the channels follow the two-level law \(1/(1+x^2)\) on average for \(x\lesssim1\); for \(x\gg1\) they share more than an isolated channel would, because product states other than the exchange partner lie within reach of the coupling (channel by channel the scatter is large); weighting each channel by the imbalance it carries in the initial state, the median \(|x|\) is \(0.035\) for \(l_{s,2}=0.50005\), \(0.35\) for \(0.5005\) and \(7.7\) for \(0.51\). At \(\lambda=10^{-3}\) they no longer do: their coupling exceeds their detuning, yet they transfer only \(40\)--\(80\%\) of their energy, because the contact has few modes (Sec.~S7).

\begin{figure}[t]
\includegraphics[width=0.75\columnwidth]{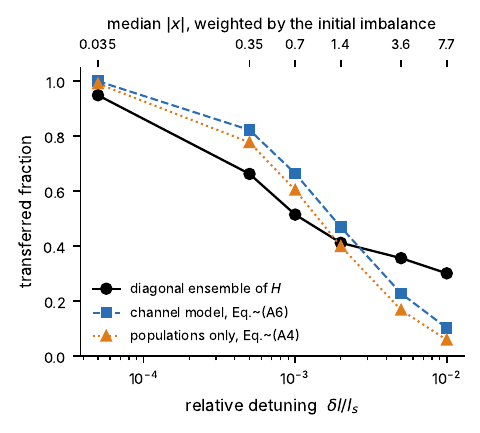}
\caption{\textbf{The channel model against the exact transfer} for detuned stadia at \(\lambda=10^{-5}\) (\(N_1=80\)). Black: diagonal ensemble of the full \(H\). Blue: Eq.~(A6) summed over channels. Orange: the same keeping only the populations, Eq.~(A4). The upper axis gives the median \(|x|=|\Delta_{ab}|/2|V_{ab}|\) of the channels, weighted by the imbalance each carries in the initial state.}
\label{fig:S_model}
\end{figure}

\paragraph{Where the channel model works.} Figure~\ref{fig:S_model} compares the model with the exact transfer. Summing Eq.~(A6) of the End Matter over the populated channels, with \(\Delta_{ab}\), \(V_{ab}\) and the initial amplitudes taken from the same basis of \(N_1=80\) levels, gives the transferred fraction without diagonalizing the coupled Hamiltonian. While the typical \(|x|\) of the populated channels is of order one or less, the model follows the diagonal ensemble of the full \(H\) within \(0.15\): for relative detunings \(5\times10^{-5}\), \(5\times10^{-4}\), \(10^{-3}\) and \(2\times10^{-3}\) it gives \(1.00\), \(0.82\), \(0.66\) and \(0.47\) against the exact \(0.95\), \(0.66\), \(0.51\) and \(0.41\). Beyond that it underestimates the transfer, down to \(0.10\) against \(0.30\) for \(10^{-2}\): once the exchange partner is far off resonance, other product states within reach of the coupling take over, and treating each channel as isolated is no longer justified. Keeping only the populations, Eq.~(A4), is always below Eq.~(A6): the coherence of the initial state adds to the transfer. The model is therefore quantitative in the region that decides the outcome, where the channels are close to resonance, and qualitative beyond it, where the temperatures stay apart in any case.

\paragraph{The direction of the flow.} The End Matter, Eq.~(A9), shows that for two thermal states every channel carries energy from hot to cold. For the coherent states used in the time evolutions, which are not thermal, the net flow is also hot to cold in all 78 time evolutions we performed (all geometries, couplings, strengths and initial states) and in all 50 ensemble predictions of Sec.~S6.

\section*{S6. OTHER INITIAL STATES}
\label{sec:S6}

\begin{figure}[t]
\includegraphics[width=0.8\columnwidth]{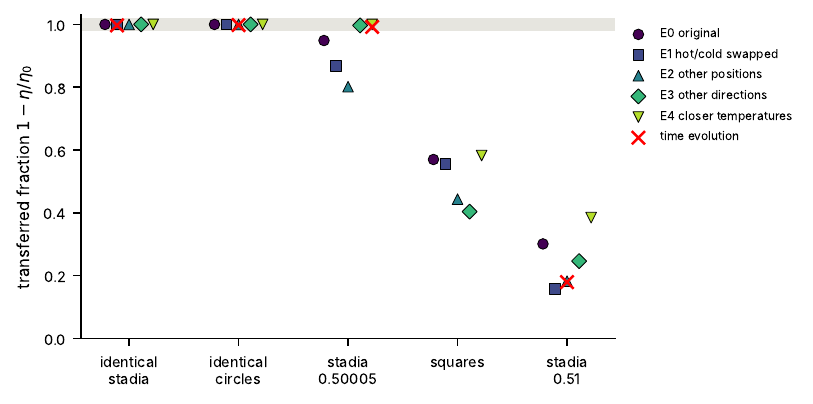}
\caption{\textbf{Other initial states.} Fraction of the imbalance transferred for five initial states (symbols) and five pairs of cavities, predicted from the eigenstates (\(\lambda=10^{-5}\)); crosses: time evolution. Shaded: equal temperatures within the fluctuations.}
\label{fig:S_states}
\end{figure}

We repeated the analysis for five initial states: the original one (E0); hot and cold interchanged, with the hot state in cavity 1 (E1); different positions of the two coherent states (E2); momenta along the axes with the same energies (E3); and closer initial temperatures (E4). Figure~\ref{fig:S_states} shows the fraction of the imbalance transferred. Identical stadia and circles equalize for every initial state, and their populations keep the memory of the initial state in all of them (total-variation distance \(\le0.04\) to the average of the two initial distributions). The squares and the stadia with \(l_{s,2}=0.51\) do not equalize for any initial state. Near the crossover, \(l_{s,2}=0.50005\), the result depends on the initial state, as expected if only some channels are Bell pairs: the residual imbalance ranges from \(9\times10^{-6}\) (E4) to \(5\times10^{-2}\) (E2). Time evolutions of four of these cases agree with the predictions: \(\Delta T/\sigma=0.11\) (identical stadia, E1), \(0.08\) (identical circles, E2), \(0.16\) (\(l_{s,2}=0.50005\), E4), and \(\eta=0.209\) against \(0.208\) predicted, \(\Delta T/\sigma=120\) (\(l_{s,2}=0.51\), E2). We also took about 800 thermal initial states per geometry, product Gibbs states of the two cavities with random phases; in all of them the heat flows from hot to cold. At \(\lambda=10^{-3}\) the results are the same for the identical cavities and the square, and the transferred fraction increases only slightly for the detuned stadia.

\section*{S7. COUPLING STRENGTH AND NUMBER OF MODES}
\label{sec:S7}

\begin{figure}[t]
\includegraphics[width=\columnwidth]{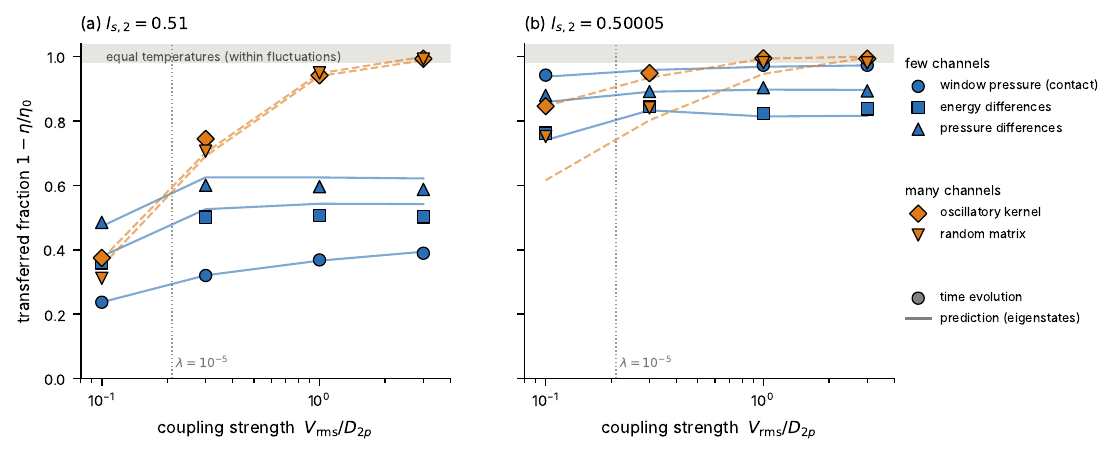}
\caption{\textbf{Coupling strength and number of modes.} Fraction of the imbalance transferred against \(V_{\rm rms}/D_{2p}\) for (a) \(l_{s,2}=0.51\) and (b) \(l_{s,2}=0.50005\). Blue: couplings with few modes [contact, Eq.~(\ref{eq:hintPSM}); \(H^{(2)}\); \(H^{(3)}\)]. Orange: couplings with many modes [\(H^{(4)}\); random matrix]. Symbols: time evolution; lines: prediction from the eigenstates. Dotted: \(\lambda=10^{-5}\) for the contact.}
\label{fig:S_coupling}
\end{figure}

Figure~\ref{fig:S_coupling} extends Fig.~3 of the main text to three other couplings and to the detuning \(l_{s,2}=0.50005\). Every point is a full time evolution (41 runs), and the lines are the predictions from the eigenstates; they agree within \(\pm0.04\) in the transferred fraction in all cases but one, the random coupling at the weakest strength and \(l_{s,2}=0.50005\) (\(0.75\) against \(0.62\)), where the time evolution and the prediction use different realizations of the random matrix. For \(l_{s,2}=0.51\) the couplings with few modes saturate: the contact transfers \(0.24\)--\(0.39\) of the imbalance, \(H^{(2)}\) about \(0.50\) and \(H^{(3)}\) about \(0.59\), with \(\Delta T/\sigma=19\)--\(90\) at all strengths. The couplings with many modes transfer \(0.94\) at \(V_{\rm rms}/D_{2p}=1\) (\(\Delta T/\sigma\simeq2\)--\(2.6\)) and \(0.99\) at \(3\) (\(\Delta T/\sigma<1\)). For the smaller detuning everything is closer to equalization: the contact transfers \(0.94\)--\(0.97\) (\(\Delta T/\sigma=1.0\)--\(2.1\)), \(H^{(2)}\) and \(H^{(3)}\) \(0.76\)--\(0.90\), and the couplings with many modes equalize from \(V_{\rm rms}/D_{2p}=1\) on. At the largest strength the interaction energy reaches \(8\%\) of the total and shifts the final temperatures; at \(V_{\rm rms}/D_{2p}\le1\) it stays below \(1\%\).

\paragraph{A physical coupling with many modes.} The random matrix can be replaced by an interaction that acts throughout the cavities. For the stadia detuned by \(2\%\) we used \(H_{\rm int}=\mathbf{p}_1\cdot\mathbf{p}_2\) and \(H_{\rm int}=\mathbf{r}_1\cdot\mathbf{r}_2\), with matrix elements computed from the interior wave functions; both act on almost all the product states (about \(6000\) independent combinations out of \(6400\)) and, unlike the contact, give the same results for \(N_1=80\) and \(120\). At \(V_{\rm rms}/D_{2p}=3\) the momentum coupling transfers \(0.95\) of the imbalance, with \(\Delta T/\sigma=2\) and an interaction energy of \(5\%\) of the total, and the position coupling \(0.78\), with \(\Delta T/\sigma=11\) and \(16\%\); the contact transfers \(0.39\) (\(0.24\) for \(N_1=120\)). At \(V_{\rm rms}/D_{2p}=10\) both reach \(0.995\) and \(\Delta T/\sigma\simeq0.2\), but the interaction energy is then \(65\)--\(90\%\) of the total: the energy is no longer the sum of those of the cavities, and their temperatures lose their meaning.

The random coupling breaks the exchange symmetry even for identical cavities. Applied to identical stadia at \(V_{\rm rms}/D_{2p}=3\), it still produces equal temperatures (\(\Delta T/\sigma=0.74\)), now by mixing many states rather than by symmetry: the diagonal elements \(\langle n|\hat T_1-\hat T_2|n\rangle\) collapse onto a narrow band around zero, as eigenstate thermalization requires. The random coupling is not a physical model of the contact; it is a control that shows what the contact lacks.

\paragraph{Eigenstate thermalization.} The diagonal elements \(\langle n|\hat T_1-\hat T_2|n\rangle\) for the contact and for the random coupling are shown in Fig.~3(a) of the main text, and their analysis in the End Matter, Eqs.~(A17)--(A18); for \(100<E_n<400\) their spread is \(0\), \(149\) and \(20\) for the three cases shown there. Equilibration in the sense of the diagonal ensemble is generic \cite{Linden2009}; what fails here for detuned cavities is the smoothness in energy required by eigenstate thermalization.

\begin{figure}[t]
\includegraphics[width=0.55\columnwidth]{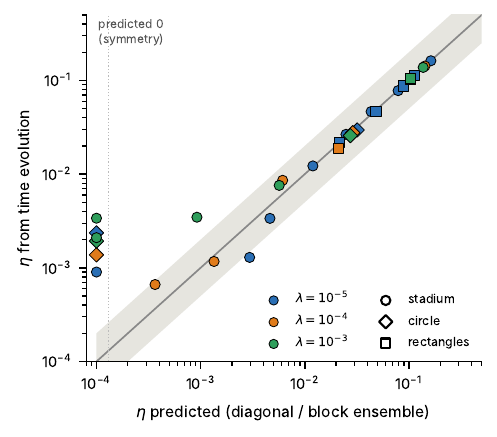}
\caption{\textbf{Time-evolved imbalance against the ensemble predictions} for the contact: the 16 clean runs at \(\lambda=10^{-5}\) and the 17 runs at \(10^{-4}\) and \(10^{-3}\). Shaded: a factor of 2 around the diagonal. Cases predicted to vanish by symmetry are plotted at the left; their time-evolved imbalance is at the level of the fluctuations.}
\label{fig:S_collapse}
\end{figure}

For the 16 cases at \(\lambda=10^{-5}\), the time-evolved imbalance agrees with Eqs.~(\ref{eq:etadiag})--(\ref{eq:etagen}) within \(8\%\) wherever it exceeds the fluctuations (rectangles within \(4\%\)). Figure~\ref{fig:S_collapse} summarizes the agreement between predictions and time evolutions for the contact at \(\lambda=10^{-5}\), \(10^{-4}\) and \(10^{-3}\): where the imbalance exceeds the fluctuations, the time-evolved value agrees with the prediction within \(10\%\).

\section*{S8. A GENERIC MODEL: TWO COUPLED RANDOM-MATRIX SUBSYSTEMS}
\label{sec:S8}

To test whether the mechanism depends on the billiards, we replaced each cavity by a random-matrix subsystem. Subsystem 1 has the spectrum \(e_i\) of a GOE matrix of size \(N=80\), the size of the billiard basis, with mean spacing \(d\); subsystem 2 is either identical or detuned, \(H_2=H_1+\delta\,d\sqrt N\,X\) with \(X\) another GOE matrix (\(\delta=1\) shifts the levels by about one spacing). The ``temperature'' of each subsystem is its energy. The initial state is a product of a cold and a hot state, Gaussian in energy (centered at \(30\%\) and \(70\%\) of the spectrum, with a width that is a fixed fraction of it) with random phases. As a contact of rank \(r\) we take
\begin{equation}
H_{\rm int}=\sum_{k=1}^{r}|w_k^{(1)}w_k^{(2)}\rangle\langle w_k^{(1)}w_k^{(2)}|,
\end{equation}
with \(w_k\) random Gaussian vectors, the same for both subsystems, as the boundary functions of a common window; for identical subsystems it respects the exchange symmetry and annihilates the antisymmetric states, as the billiard contact does. We compare it with a GOE coupling of full rank, and measure the strength by \(V_{\rm rms}/D_{2p}\) as in Sec.~S2. The transferred fraction is computed from the diagonal ensemble and averaged over 20 realizations.

\begin{figure}[t]
\includegraphics[width=\columnwidth]{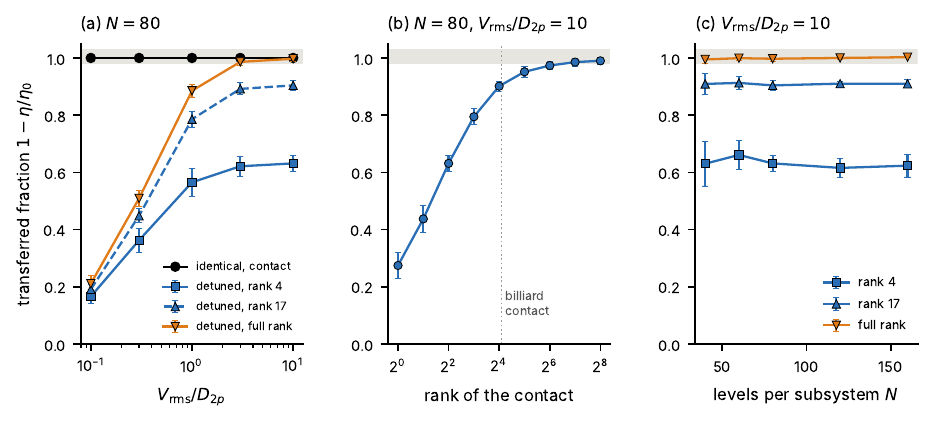}
\caption{\textbf{The mechanism in two random-matrix subsystems} (\(N=80\) levels each unless stated; mean and spread over 20 realizations). (a) Fraction of the imbalance transferred against the coupling strength: identical subsystems with a symmetric contact (black), and detuned subsystems (\(\delta=1\)) with contacts of rank 4 and 17 (blue) and a full-rank GOE coupling (orange). (b) Transfer at strong coupling against the rank of the contact; dotted: rank of the billiard contact. (c) Transfer at strong coupling against the number of levels per subsystem (20 realizations for \(N\le80\), 10 for \(N=120\), 3 for \(N=160\)). Shaded: complete transfer.}
\label{fig:S_rmt}
\end{figure}

Figure~\ref{fig:S_rmt} reproduces the three results of the main text. (i) For identical subsystems the symmetric contact equalizes the energies exactly at every strength, while the full-rank GOE coupling, which breaks the symmetry, does so only at strong coupling. (ii) For detuned subsystems the contact saturates: from \(V_{\rm rms}/D_{2p}=1\) to \(10\) the transfer only goes from \(0.57\) to \(0.63\) for \(r=4\) and from \(0.79\) to \(0.90\) for \(r=17\), while the full-rank coupling reaches complete transfer. (iii) The upper limit rises with the rank of the contact [Fig.~\ref{fig:S_rmt}(b)], from \(0.27\) for \(r=1\) to \(0.90\) for \(r=17\) and \(0.99\) for \(r\ge128\); it is the same for detunings \(\delta=0.2\) and \(1\).

The upper limit does not depend on the size of the subsystems [Fig.~\ref{fig:S_rmt}(c)]: for \(N=40\) to \(160\), with a number of populated levels growing proportionally to \(N\), it stays at \(0.91\) for \(r=17\) and at \(0.62\)--\(0.66\) for \(r=4\). The bottleneck of a low-rank contact is therefore not a finite-size effect. The upper limit of the billiard contact (\(0.43\) for \(r=17\)) is lower than that of random vectors of the same rank, because its boundary functions are smooth in energy and correlated from level to level; the existence of an upper limit below one, set by the rank, is generic.

\begin{figure}[t]
\includegraphics[width=0.62\columnwidth]{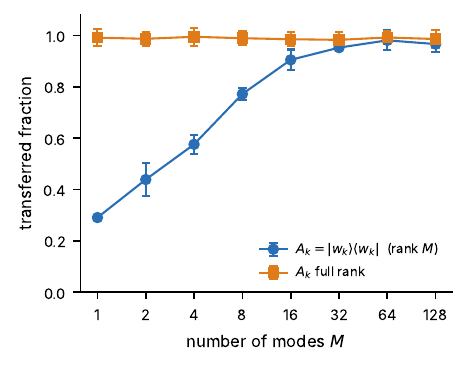}
\caption{\textbf{Rank, not number of modes.} Transferred fraction against the number of modes \(M\) of the contact \(H_{\rm int}=\sum_k A_k\otimes A_k\), for detuned random-matrix subsystems (\(N=48\), \(\delta=1\), \(V_{\rm rms}/D_{2p}=3\), 3 realizations). Blue: \(A_k\) projectors, so that the rank in the product space is \(M\). Orange: \(A_k\) of full rank, with the same number of modes.}
\label{fig:S_rank}
\end{figure}

\paragraph{Rank or number of modes?} The two can be separated in this model. Writing the contact as \(H_{\rm int}=\sum_{k=1}^{M}A_k\otimes A_k\), the number of modes is \(M\), while its rank in the product space depends on the \(A_k\): for the projectors used above the rank is \(M\), whereas for full-rank \(A_k\) it is of order \(N^2\) with the same \(M\). With \(M=4\) and detuned subsystems (\(N=80\)), the projector contact transfers \(0.58\) to \(0.64\) from \(V_{\rm rms}/D_{2p}=1\) to \(10\), while with full-rank \(A_k\) it transfers \(0.88\) to \(0.99\), as a GOE coupling of full rank does (\(0.88\) to \(0.99\)); the same holds for \(N=48\). Against the number of modes [Fig.~\ref{fig:S_rank}], the projector contact rises slowly, \(0.29\), \(0.58\), \(0.91\) and \(0.98\) for \(M=1\), \(4\), \(16\) and \(64\), while with full-rank \(A_k\) a single mode already gives complete transfer. For identical subsystems both give complete transfer at every rank and every strength, to \(10^{-12}\), so the symmetry result does not depend on it. What limits the transfer is therefore the rank in the product space, not the number of modes. For one particle per cavity the two coincide, because the pressure at a point, \(\hat P(y)=|\chi_y\rangle\langle\chi_y|\), is a projector; for many particles per cavity the same few modes are built from one-body operators of high rank, and the argument of Appendix E does not apply.

\section*{S9. ULTRAVIOLET CUTOFF OF THE CONTACT}
\label{sec:S9}

\paragraph{Why a cutoff is needed.} The pressure \(|\partial_n\psi|^2\) grows as \(k^2\), so the matrix elements of \(H_{\rm int}\) grow with the levels involved: the largest eigenvalue of \(H_{\rm int}\) is \(s_1=1.3\times10^8\), \(4.8\times10^8\) and \(1.3\times10^9\) for \(N_1=80\), \(120\) and \(160\) (stadium). At \(\lambda=10^{-5}\) the strongest directions, built from the highest levels kept, already satisfy \(\lambda s_j\gg\sigma_E\); they behave as the protected directions of Eq.~(A16) and reduce the effective coupling within the energy shell. As for any contact interaction, \(\lambda\) alone is therefore not a physical parameter: the basis acts as the ultraviolet cutoff and is part of the model. At fixed \(\lambda=10^{-5}\) the transfer decreases with \(N_1\) (Table~\ref{tab:uv}); at \(\lambda=10^{-7}\), where \(\lambda s_1\) is smaller, it changes by less than \(5\%\) from \(N_1=120\) to \(160\).

\paragraph{A smooth cutoff.} A real wall cannot follow pressure fluctuations of arbitrarily short wavelength. We model this by \(\partial_n\phi_a\to\partial_n\phi_a\,e^{-\varepsilon_a/2E_c}\) in both cavities. The factor preserves the exchange symmetry and, since its product over a channel depends only on the total energy, the dark states of the rectangles. With \(E_c=300\) the results converge in \(N_1\) to within \(4\%\) (Table~\ref{tab:uv}). Compared at the same \(V_{\rm rms}/D_{2p}=0.21\) with the model of the main text, identical stadia still equalize exactly; the stadia detuned by \(2\%\) transfer \(0.39\) instead of \(0.30\), with \(\Delta T/\sigma=53\) instead of \(82\); and \(l_{s,2}=0.50005\) transfers \(0.94\) instead of \(0.95\), with \(\Delta T/\sigma\simeq2\) in both. The conclusions are unchanged. The strong-coupling upper limit, in contrast, depends on the cutoff: for \(l_{s,2}=0.51\) and \(N_1=120\) it is \(0.76\), \(0.58\) and \(0.48\) for \(E_c=300\), \(500\) and \(800\), and \(0.25\) without the smooth cutoff, always well below one (\(0.97\)--\(0.98\) for \(l_{s,2}=0.50005\)). The rank within the energy shell, \(r_{\rm shell}=11\), is the same in all cases. Against the coupling strength [open circles in Fig.~3(b) of the main text], the transfer for \(l_{s,2}=0.51\) with \(E_c=300\) and \(N_1=120\) rises from \(0.25\) at \(V_{\rm rms}/D_{2p}=0.1\) to \(0.70\) at \(3\) and \(0.74\) at \(10\), approaching its upper limit, while \(\Delta T/\sigma\) falls from \(89\) to \(15\); with \(E_c=500\) it saturates at \(0.64\). With the cutoff of the main text the same range gives \(0.24\) to \(0.41\) and \(\Delta T/\sigma\) from \(100\) to \(57\). In every case the detuned stadia stay far from equal temperatures, whereas for \(l_{s,2}=0.50005\) the gap reaches the fluctuations, \(\Delta T/\sigma\simeq1\), for \(V_{\rm rms}/D_{2p}\gtrsim1\).

\paragraph{Reweighting the levels.} The growth of the matrix elements with the energy of the levels involved, noted above, is not a choice of the model. For a normalized Dirichlet eigenfunction \(\oint|\partial_n\phi|^2(\mathbf{r}\cdot\mathbf{n})\,ds=2k^2\), so \(\partial_n\phi\propto k\) and \(\langle a|\hat P|a\rangle\propto\varepsilon_a\): this is the relation between pressure and energy of the two-dimensional ideal gas on which the definition of the temperature rests \cite{Rivas2025}. A hotter particle pushes harder on the wall; what is the same throughout the spectrum is \(\lambda\), the response of the wall to a given force. One may nevertheless ask what would change if the growth were removed by hand. Any positive weight applied to the boundary functions of both cavities, \(\partial_n\phi_a\to g_a\,\partial_n\phi_a\), acts on the coupling as \(H_{\rm int}\to(D\otimes D)\,H_{\rm int}\,(D\otimes D)\), with \(D=\mathrm{diag}(g_a)\) invertible and diagonal in the product basis. It therefore leaves exactly unchanged (i) the rank of the contact, in the whole space and within any energy shell, since the shell projector is diagonal in the same basis; (ii) the number of dark states of each degenerate group, \(m_g\) minus the rank of its block; and (iii) the exchange symmetry of identical cavities, so that they still equalize exactly. Which combinations are dark does change. Moreover, the two states of a channel are built from the same pair of levels and so carry the same factor \(g_a^2g_b^2\): the detuning \(\Delta_{ab}\) is untouched and only \(x=\Delta_{ab}/2V_{ab}\) is rescaled, channel by channel, which is a redefinition of \(\lambda\) -- and Sec.~S7 shows that changing \(\lambda\) over three orders of magnitude, up to the strong-coupling limit of Eq.~(A16), does not equalize the detuned cavities. The smooth cutoff above is one such weight, \(g_a=e^{-\varepsilon_a/2E_c}\); the weight that cancels the growth of the pressure altogether is \(g_a=1/k_a\), and for it \(g_a^2g_b^2=(\varepsilon_a\varepsilon_b)^{-1}\) varies within a factor of two across \(90\%\) of the channels populated by the initial state, and by a factor of eight between the extremes (40 most populated channels of the stadia detuned by \(2\%\)), so that it amounts to little more than a channel-dependent redefinition of \(\lambda\). The energy profile of the contact is thus not what limits the transfer: among the couplings of Sec.~S2, \(H^{(2)}\) and \(H^{(3)}\) grow steeply with the energy and do not equalize the detuned cavities, whereas \(H^{(4)}\), whose matrix elements have the same size throughout the spectrum, does -- because it acts on many directions, not because it is flat in energy.

\paragraph{Width of the window.} The number of modes of the window sets the upper limit. Keeping the energy shell fixed and centering the window on the one of the main text, its width \(w\) changes \(r_{\rm shell}\) and the upper limit of Eq.~(A16) together: for \(l_{s,2}=0.51\) and \(N_1=80\), \(w/w_0=0.15\), \(0.3\), \(0.5\), \(0.75\), \(1\) and the whole common wall give \(r_{\rm shell}=6\), \(7\), \(10\), \(11\), \(11\) and \(12\), and upper limits \(0.27\), \(0.33\), \(0.38\), \(0.42\), \(0.43\) and \(0.48\) (the same trend, shifted down, for \(N_1=120\)). For product states of energy up to \(E\) the profiles oscillate with wave numbers up to \(K=\sqrt{2E}\), and the estimate \(wK/\pi\), the number of transverse modes of a constriction of width \(w\), gives \(8.8\) for the window of the main text and \(E=E_0+2\sigma_E\), against \(r_{\rm shell}=10\) or \(11\) depending on the criterion (\(99.99\%\) of the trace, or eigenvalues above \(10^{-8}\) of the largest); for narrow windows the count exceeds the estimate by a few modes. The shell used for \(r_{\rm shell}\) is \(|E_{ab}-E_0|<2\sigma_E\); \(V_{\rm rms}\) and \(D_{2p}\) are computed within \(1\sigma_E\).

\paragraph{Tolerance at strong coupling.} A stronger contact widens the range of detunings that equalize, but only up to a point. With the smooth cutoff (\(E_c=300\), \(N_1=120\)), the strong-coupling limit of Eq.~(A16) gives, for relative detunings \(10^{-4}\), \(10^{-3}\), \(2\times10^{-3}\), \(4\times10^{-3}\), \(10^{-2}\) and \(2\times10^{-2}\), transferred fractions \(0.98\), \(0.90\), \(0.86\), \(0.82\), \(0.77\) and \(0.76\), and \(\Delta T/\sigma=0.9\), \(4\), \(6\), \(9\), \(12\) and \(13\); at \(V_{\rm rms}/D_{2p}=3\) the gaps are \(1.0\), \(6\), \(9\), \(12\), \(15\) and \(18\). With the cutoff of the main text the same limit gives \(\Delta T/\sigma=0.8\), \(10\), \(18\), \(26\), \(41\) and \(53\). Only the smallest detuning equalizes.

\paragraph{Energy of the initial state.} Scaling the momenta of the initial state by \(\sqrt S\), and the cutoff with it (\(E_c=300\,S\)), the number of modes grows slowly, as the estimate predicts: for \(S=0.5\), \(1\), \(1.5\) and \(2\), \(wK/\pi=8.0\), \(9.0\), \(9.8\) and \(10.4\), and \(r_{\rm shell}=10\), \(11\), \(12\) and \(13\), while the number of product states in the shell grows from \(1463\) to \(4702\). For \(l_{s,2}=0.51\) the strong-coupling limit of the transfer falls from \(0.95\) to \(0.76\), \(0.59\) and \(0.55\), and \(\Delta T/\sigma\) in that limit rises from \(1.2\) to \(13\), \(40\) and \(61\). Part of this fall is due to the cutoff, which here grows with the energy: keeping \(E_c=300\) at \(S=2\) the limit is \(0.73\), close to the \(0.76\) of \(S=1\), but the gap still doubles, \(\Delta T/\sigma=29\). Taking the whole common wall as the window at \(S=2\) raises the limit only from \(0.55\) to \(0.59\), with \(r_{\rm shell}=14\). The tolerance to detuning shrinks with energy: for \(l_{s,2}=0.50005\) the strong-coupling limit gives \(\Delta T/\sigma=0.5\), \(0.9\), \(2.0\) and \(3.4\) for the four values of \(S\), so that a detuning of \(10^{-4}\) no longer equalizes at the highest energy; the same holds with the cutoff kept fixed at \(E_c=300\) (\(\Delta T/\sigma=3.2\) at \(V_{\rm rms}/D_{2p}=3\) and \(2.7\) in the strong-coupling limit, for \(S=2\)). At fixed window the bottleneck therefore does not weaken with energy: the modes grow only as the wave number, while the detunings of the channels in units of the level spacing grow faster.

\paragraph{Interaction energy.} The temperatures used throughout require the energy to be the sum of those of the cavities. This holds for the contact at all the strengths considered: in the initial state of the stadia detuned by \(2\%\), \(\lambda\langle H_{\rm int}\rangle\) is \(0.01\%\), \(0.13\%\) and \(0.45\%\) of the total energy at \(V_{\rm rms}/D_{2p}=0.21\), \(3\) and \(10\) with the cutoff of the main text, and \(0.05\%\), \(0.7\%\) and \(2.3\%\) with the smooth cutoff; the initial state has a weight below \(10^{-3}\) on the combinations on which the contact acts. A contact can thus be strong for its few modes and still negligible in energy, in contrast with the couplings of many modes of Sec.~S7.

\paragraph{A nonlocal wall.} The contact of the main text is local: the pressure of one particle at a point of the window acts on the other at the same point. A wall with tension would respond nonlocally. We tested \(H_{\rm int}=\int_w\!\int_w P_1(y)\,G_\ell(y-y')\,P_2(y')\,dy\,dy'\) with a normalized Gaussian kernel of range \(\ell\), from \(\ell=0\) (the contact) to \(\ell=\infty\) (the product of the total pressures on the window), for the stadia detuned by \(2\%\) with the smooth cutoff and \(N_1=80\). The number of modes grows with the range, \(r_{\rm shell}=11\), \(35\), \(52\) and \(52\) for \(\ell=0\), \(0.02\), \(0.1\) and \(\infty\), but the transfer barely does: at \(V_{\rm rms}/D_{2p}=3\) it is \(0.73\), \(0.72\), \(0.75\) and \(0.79\), with \(\Delta T/\sigma=15\), \(15\), \(13\) and \(11\); at \(V_{\rm rms}/D_{2p}=10\), \(0.76\) to \(0.83\). The kernel changes only how the pressure profiles overlap, not the profiles themselves, and fifty modes remain few compared with the thousands of states of the energy shell; what limits the transfer is the resonance of the channels. Identical stadia equalize exactly with every kernel. Full time evolutions with the smooth cutoff confirm the predictions from the eigenstates: for the detuned stadia at \(V_{\rm rms}/D_{2p}=0.1\) to \(3\) the transferred fractions agree within \(0.01\) and \(\Delta T/\sigma\) within \(15\%\) [symbols in Fig.~3(b),(c) of the main text], and at \(0.21\) the five pairs of Fig.~1 give \(1.000\) (identical stadia), \(0.41\) (\(l_{s,2}=0.51\), \(\Delta T/\sigma=43\)), \(0.96\) (\(0.50005\), \(1.6\)), \(0.99\) (circles) and \(0.57\) (squares, \(13\)), the same picture as with the cutoff of the main text. The interaction energy of the contact in the diagonal ensemble is \(0.3\%\) of the total at \(V_{\rm rms}/D_{2p}=3\) and \(1.7\%\) at \(10\) (smooth cutoff).

\begin{table}[h]
\caption{Dependence on the cutoff, stadia with \(l_{s,1}=0.5\). Transferred fraction \(1-\eta/\eta_0\) from the diagonal ensemble; upper limit from Eq.~(A16).}
\label{tab:uv}
\begin{ruledtabular}
\begin{tabular}{lccc}
 & \(N_1=80\) & \(120\) & \(160\)\\
\hline
\multicolumn{4}{l}{Basis as cutoff, \(\lambda=10^{-5}\) (\(V_{\rm rms}/D_{2p}\simeq0.21\))}\\
\quad \(l_{s,2}=0.51\) & 0.30 & 0.21 & 0.14\\
\quad \(l_{s,2}=0.50005\) & 0.95 & 0.91 & 0.84\\
\quad upper limit, \(l_{s,2}=0.51\) & 0.43 & 0.25 & 0.16\\
\multicolumn{4}{l}{Smooth cutoff \(E_c=300\), \(V_{\rm rms}/D_{2p}=0.21\)}\\
\quad \(l_{s,2}=0.51\) & 0.41 & 0.40 & 0.39\\
\quad \(l_{s,2}=0.50005\) & 0.96 & 0.94 & 0.94\\
\quad upper limit, \(l_{s,2}=0.51\) & 0.78 & 0.76 & 0.76\\
\hline
rank \(r\) (\(s_j/s_1>10^{-8}\)) & 17 & 20 & 23\\
rank in the energy shell \(r_{\rm shell}\) & 11 & 11 & 11\\
\end{tabular}
\end{ruledtabular}
\end{table}


\begin{thebibliography}{99}
\bibitem{Brown2001} H. R. Brown and J. Uffink, Stud. Hist. Phil. Mod. Phys. \textbf{32}, 525 (2001).
\bibitem{Gemmer2009} J. Gemmer, M. Michel, and G. Mahler, \textit{Quantum Thermodynamics} (Springer, 2009).
\bibitem{Gogolin2016} C. Gogolin and J. Eisert, Rep. Prog. Phys. \textbf{79}, 056001 (2016).
\bibitem{Partovi2008} M. H. Partovi, Phys. Rev. E \textbf{77}, 021110 (2008).
\bibitem{Jennings2010} D. Jennings and T. Rudolph, Phys. Rev. E \textbf{81}, 061130 (2010).
\bibitem{Micadei2019} K. Micadei, J. P. S. Peterson, A. M. Souza, R. S. Sarthour, I. S. Oliveira, G. T. Landi, T. B. Batalh\~ao, R. M. Serra, and E. Lutz, Nat. Commun. \textbf{10}, 2456 (2019).
\bibitem{Bunimovich1979} L. A. Bunimovich, Commun. Math. Phys. \textbf{65}, 295 (1979).
\bibitem{Shnirelman1974} A. I. Shnirelman, Usp. Mat. Nauk \textbf{29}, 181 (1974).
\bibitem{Bohigas1984} O. Bohigas, M. J. Giannoni, and C. Schmit, Phys. Rev. Lett. \textbf{52}, 1 (1984).
\bibitem{Deutsch1991} J. M. Deutsch, Phys. Rev. A \textbf{43}, 2046 (1991).
\bibitem{Srednicki1994} M. Srednicki, Phys. Rev. E \textbf{50}, 888 (1994).
\bibitem{Rigol2008} M. Rigol, V. Dunjko, and M. Olshanii, Nature \textbf{452}, 854 (2008).
\bibitem{Dalessio2016} L. D'Alessio, Y. Kafri, A. Polkovnikov, and M. Rigol, Adv. Phys. \textbf{65}, 239 (2016).
\bibitem{Borgonovi2016} F. Borgonovi, F. M. Izrailev, L. F. Santos, and V. G. Zelevinsky, Phys. Rep. \textbf{626}, 1 (2016).
\bibitem{Rigol2007} M. Rigol, V. Dunjko, V. Yurovsky, and M. Olshanii, Phys. Rev. Lett. \textbf{98}, 050405 (2007).
\bibitem{Miller1999} P. A. Miller and S. Sarkar, Phys. Rev. E \textbf{60}, 1542 (1999).
\bibitem{Kaufman2016} A. M. Kaufman, M. E. Tai, A. Lukin, M. Rispoli, R. Schittko, P. M. Preiss, and M. Greiner, Science \textbf{353}, 794 (2016).
\bibitem{Bergamasco2017} P. D. Bergamasco, G. G. Carlo, and A. M. F. Rivas, Phys. Rev. E \textbf{96}, 062144 (2017).
\bibitem{Bergamasco2019} P. D. Bergamasco, G. G. Carlo, and A. M. F. Rivas, Phys. Rev. Res. \textbf{1}, 033044 (2019).
\bibitem{Goldstein2006} S. Goldstein, J. L. Lebowitz, R. Tumulka, and N. Zangh\`i, Phys. Rev. Lett. \textbf{96}, 050403 (2006).
\bibitem{Popescu2006} S. Popescu, A. J. Short, and A. Winter, Nat. Phys. \textbf{2}, 754 (2006).
\bibitem{Ponomarev2011} A. V. Ponomarev, S. Denisov, and P. H\"anggi, Phys. Rev. Lett. \textbf{106}, 010405 (2011).
\bibitem{Ponomarev2012} A. V. Ponomarev, S. Denisov, J. Gemmer, and P. H\"anggi, EPL \textbf{98}, 40011 (2012).
\bibitem{Lakshminarayan2016} A. Lakshminarayan, S. C. L. Srivastava, R. Ketzmerick, A. B\"acker, and S. Tomsovic, Phys. Rev. E \textbf{94}, 010205(R) (2016).
\bibitem{Tomsovic2018} S. Tomsovic, A. Lakshminarayan, S. C. L. Srivastava, and A. B\"acker, Phys. Rev. E \textbf{98}, 032209 (2018).
\bibitem{Pulikkottil2023} J. J. Pulikkottil, A. Lakshminarayan, S. C. L. Srivastava, M. F. I. Kieler, A. B\"acker, and S. Tomsovic, Phys. Rev. E \textbf{107}, 024124 (2023).
\bibitem{Stockmann1990} H.-J. St\"ockmann and J. Stein, Phys. Rev. Lett. \textbf{64}, 2215 (1990).
\bibitem{Marcus1992} C. M. Marcus, A. J. Rimberg, R. M. Westervelt, P. F. Hopkins, and A. C. Gossard, Phys. Rev. Lett. \textbf{69}, 506 (1992).
\bibitem{Rivas2025} A. M. F. Rivas, E. G. Vergini, L. Ermann, and G. G. Carlo, Phys. Rev. E \textbf{112}, 014223 (2025).
\bibitem{Barnett2000} A. Barnett, D. Cohen, and E. J. Heller, Phys. Rev. Lett. \textbf{85}, 1412 (2000).
\bibitem{Vergini1995} E. Vergini and M. Saraceno, Phys. Rev. E \textbf{52}, 2204 (1995).
\bibitem{Lieb1999} E. H. Lieb, Physica A \textbf{263}, 491 (1999).
\bibitem{Gruber1999} Ch. Gruber, Eur. J. Phys. \textbf{20}, 259 (1999).
\bibitem{Kestemont2000} E. Kestemont, C. Van den Broeck, and M. Malek Mansour, Europhys. Lett. \textbf{49}, 143 (2000).
\bibitem{CasasVazquez2003} J. Casas-V\'azquez and D. Jou, Rep. Prog. Phys. \textbf{66}, 1937 (2003).
\bibitem{Puglisi2017} A. Puglisi, A. Sarracino, and A. Vulpiani, Phys. Rep. \textbf{709}, 1 (2017).
\bibitem{Rego1998} L. G. C. Rego and G. Kirczenow, Phys. Rev. Lett. \textbf{81}, 232 (1998).
\bibitem{Schwab2000} K. Schwab, E. A. Henriksen, J. M. Worlock, and M. L. Roukes, Nature \textbf{404}, 974 (2000).
\bibitem{Facchi2002} P. Facchi and S. Pascazio, Phys. Rev. Lett. \textbf{89}, 080401 (2002).
\bibitem{Krinner2015} S. Krinner, D. Stadler, D. Husmann, J.-P. Brantut, and T. Esslinger, Nature \textbf{517}, 64 (2015).
\bibitem{Brantut2013} J.-P. Brantut, C. Grenier, J. Meineke, D. Stadler, S. Krinner, C. Kollath, T. Esslinger, and A. Georges, Science \textbf{342}, 713 (2013).
\bibitem{Zhang2026} X. Zhang, J. Che, and B. Dietz, Phys. Rev. E \textbf{113}, 024214 (2026).
\bibitem{Page1993} D. N. Page, Phys. Rev. Lett. \textbf{71}, 1291 (1993).
\bibitem{Oganesyan2007} V. Oganesyan and D. A. Huse, Phys. Rev. B \textbf{75}, 155111 (2007).
\bibitem{Atas2013} Y. Y. Atas, E. Bogomolny, O. Giraud, and G. Roux, Phys. Rev. Lett. \textbf{110}, 084101 (2013).
\bibitem{Linden2009} N. Linden, S. Popescu, A. J. Short, and A. Winter, Phys. Rev. E \textbf{79}, 061103 (2009).
\end{thebibliography}
\end{document}